\documentclass[sigconf]{acmart} 
\AtBeginDocument{%
  }

\copyrightyear{2026}
\acmYear{2026}
\setcopyright{cc}
\setcctype{by}
\acmConference[CHI '26]{Proceedings of the 2026 CHI Conference on Human Factors in Computing Systems}{April 13--17, 2026}{Barcelona, Spain}
\acmBooktitle{Proceedings of the 2026 CHI Conference on Human Factors in Computing Systems (CHI '26), April 13--17, 2026, Barcelona, Spain}
\acmPrice{}
\acmDOI{10.1145/3772318.3790984}
\acmISBN{979-8-4007-2278-3/2026/04}

\usepackage{enumitem}
\usepackage{longtable}
\usepackage{booktabs}
\usepackage{tabularx}

\AtBeginDocument{
  }

\newcommand{\add}[1]{\textcolor{black}{#1}}

\begin{document}

\author{Qing Xiao}
\affiliation{
  \institution{Human-Computer Interaction Institute, Carnegie Mellon University}
  \city{Pittsburgh}
  \state{Pennsylvania}
  \country{USA}
}
\email{qingx@cs.cmu.edu}

\author{Qing Hu}
\affiliation{
  \institution{Human-Computer Interaction Institute, Carnegie Mellon University}
  \city{Pittsburgh}
  \state{Pennsylvania}
  \country{USA}
}
\email{dianehu@andrew.cmu.edu}

\author{Jingjia Xiao}
\affiliation{
  \institution{Department of Sociology, University of California San Diego}
  \city{La Jolla}
  \state{California}
  \country{USA}
}
\email{jix082@ucsd.edu}

\author{Hancheng Cao}
\affiliation{
  \institution{Goizueta Business School, Emory University}
  \city{Atlanta}
  \state{Georgia}
  \country{USA}
}
\email{hancheng.cao@emory.edu}

    \author{Hong Shen}
\affiliation{
  \institution{Human-Computer Interaction Institute, Carnegie Mellon University}
  \city{Pittsburgh}
  \state{Pennsylvania}
  \country{USA}
}
\email{hongs@cs.cmu.edu}

\title[Experiences, Challenges, and Opportunities of Coordinating GenAI into Collaborative Newswork]{Can GenAI Move from Individual Use to Collaborative Work? Experiences, Challenges, and Opportunities of Coordinating GenAI into Collaborative Newswork}

\begin{abstract}
Generative AI (GenAI) is reshaping work, but adoption remains largely individual and experimental rather than coordinated into collaborative work. Whether GenAI can move from individual use to collaborative work is a critical question for future organizations. Journalism offers a compelling site to examine this shift: individual journalists have already been disrupted by GenAI tools; yet newswork is inherently collaborative relying on shared norms and coordinated workflows. We conducted 27 interviews with newsroom managers, editors and front-line journalists in China. We found that journalists frequently used GenAI to support daily tasks, but value alignment was safeguarded mainly through individual discretion. At the organizational level, GenAI use remained disconnected from team workflows, hindered by structural barriers and cultural reluctance to share practices. These findings underscore the gap between individual and collaborative work, pointing to the need to account for organizational structures, cultural norms, and workflow when coordinating GenAI for collaborative work.
\end{abstract}

\ccsdesc[500]{Human-centered computing~Empirical studies in collaborative and social computing}
\ccsdesc[300]{Human-centered computing~Empirical studies in HCI}

\keywords{GenAI, journalism, organizations, teamwork, collaborative work, human-AI collaboration}

\maketitle
\section{Introduction}

\add{Existing discussions of GenAI adoption often assume that organizations play an active, top-down role in deciding when and how to integrate GenAI \cite{seifdar2025strategic,xiao2025might}. However, in many real-world settings, GenAI enters organizations not through coordinated institutional adoption but through dispersed, informal, and individualized practices \cite{waters2025shadow,wu2024journalists}. Some individuals are already using GenAI inside organizations, often invisibly, creating pressures, ambiguities, and expectations for future collaborative coordination. Accordingly, our study asks how the widespread yet individualized GenAI use already occurring inside organizations creates new pressures for collaborative coordination and team workflow, and how organizations might respond by meaningfully coordinate GenAI into collaborative work under these conditions.}

In this paper, we use the newsroom as our primary empirical site to examine how individualized GenAI usage interacts with organizational collaboration. News production is rarely the effort of a single individual: front-line journalists gather information, draft stories, and exchange leads with peers; their work is then reviewed, edited, and sometimes rewritten by editors; and senior managers provide strategic oversight, shape policies, and liaise with external stakeholders \cite{duffy2021out,xiao2025might,tuchman1973making}. Collaboration in newsrooms therefore spans both inter-role and intra-role interactions, anchored in shared norms and collective values such as editorial responsibility and mutual trust \cite{tuchman1973making,carlson2009dueling,robinson2011journalism}. 

While these roles structure collaboration in conventional newswork, the rise of GenAI introduces new uncertainties about how such collaboration could unfold. Recently, on the one hand, many individual journalists are experimenting with GenAI in tentative, ad hoc ways \cite{Liang2025widespread}, often relying on personal judgment to ensure that its use remains consistent with professional norms \cite{wu2024journalists}. On the other hand, major news organizations have started to articulate institutional responses to the rise of GenAI, for example, China Media Group (CMG) is developing proprietary GenAI systems for its news production \cite{chinaxinwen2024aimedia,lang2024qianqiu_shisong}, while The British Broadcasting Corporation (BBC) has issued formal usage guidelines for individual journalists~\cite{bbc_ai_guidance_2024}. \add{However, these organizational responses do not yet address a crucial issue: how the already-prevalent individual uses of GenAI could be coordinated or aligned within collaborative newswork}: how teams of journalists and editors might coordinate their GenAI use and maintain shared editorial standards in future collaborations. 

\add{This gap between widespread individual use of GenAI and the absence of collaborative structures is the central problem this study addresses.} To understand why this missing layer of collaboration matters, we turn to the normative features of journalism that make it a particularly high-stakes and revealing site for studying GenAI use in collaborative work. First, as a content-intensive profession, journalism is especially exposed to GenAI’s disruptions: core newsroom tasks such as brainstorming, reporting, and editing directly overlap with GenAI’s generative capabilities in text production, compelling journalists to negotiate how these systems intervene in their everyday collaborations \cite{cools2024uses}. Second, newswork is inherently collaborative, structured through collective editorial routines and shared news values \cite{schudson2001objectivity,deuze2005journalism}. GenAI therefore does not simply alter individual tasks but also potentially reshapes the collaborations and mutual trust on which newsroom collaboration depends. Third, the public visibility of journalism subjects GenAI adoption to heightened scrutiny, both inside and outside the newsroom. This visibility amplifies the stakes of collaborative decisions about when and how to use GenAI, since any misstep is not only organizationally consequential but also publicly accountable \cite{petre2021all,chadha2016re,nishal2025values,moller2024designing,komatsu2020ai}.

Yet despite journalism’s collaborative nature, much of the existing research on GenAI use has remained focused at the individual level. Early research on GenAI in newsrooms has largely focused on task-level affordances and risks for individuals. Scholars have highlighted how AI tools can accelerate content creation and support editorial work, while raising concerns about factual errors, bias amplification, content homogenization, and the erosion of core journalistic skills \cite{noh2025biassist,wang2025media,woodruff2024how,siitonen2024mapping}. More recent work has turned attention to value alignment and professional norms, showing that individual journalists adopt GenAI selectively and cautiously, favoring uses that preserve editorial credibility \cite{nishal2025values,tseng2025ownership,xiao2025might,wu2024journalists}. However, a key gap remains: we know little about how GenAI becomes, or fails to become, coordinated into collaborative newsroom newswork. Here, we refer not to sporadic or experimental uses of GenAI, but to its potential for sustained integration into everyday collaborative newswork, where its presence is taken for granted, embedded in ongoing teamwork, and coordinated with shared professional norms.

Therefore, our study asks:
\begin{itemize}
    \item \add{\textbf{RQ1:} How does the existing individual use of GenAI affect collaborative work in newsrooms?}
    \item \add{\textbf{RQ2:} What challenges do journalists face in coordinating GenAI into their collaborative newswork?}
    \item \add{\textbf{RQ3:} How do journalists envision future organizational mechanisms and system design that could better coordinate GenAI use in collaborative newswork?}
\end{itemize}

To answer these questions, we conducted an in-depth qualitative study of Chinese newsrooms. China is a fertile site for studying GenAI's impact in newsroom collaborations because Chinese newsrooms operate under both rapid technological transformation and strict professional expectations. Journalists navigate a dual pressure: leveraging cutting-edge GenAI tools to enhance innovation under governmental expectations, while upholding the editorial authority that underpins their professional norms \cite{xiao2025might}. This dual pressure magnifies the challenges of coordinating GenAI into collaborative newswork and makes teamwork negotiation around GenAI especially visible.

Our study draws on 27 semi-structured interviews with senior newsroom managers, editors, and front-line journalists in China who had participated in early GenAI adoption in their newsrooms. Our findings showed that GenAI is already widely used at the individual level in Chinese newsrooms, but such use remains highly personalized, informal, and largely disconnected from collaborative newswork (\textbf{\textit{RQ1}}). Then, we found that structural fragmentation at the organizational level and cultural reluctance to share practices have prevented GenAI from being coordinated into collaborative newswork. \add{Current Chinese newsrooms have invested heavily in developing internal GenAI systems, yet have devoted far less attention to coordinating GenAI use within collaborative newswork} (\textbf{\textit{RQ2}}). Journalists also envision future coordination of GenAI into collaborative newswork as requiring a rethinking of organizational structures, cultural norms, and workflow (\textbf{\textit{RQ3}}).

This study makes three key contributions to HCI and CSCW community:
\begin{itemize}
    \item First, we provide empirical evidence showing that GenAI use in newsrooms is currently individualized rather than collaborative. We highlight the gap between personal experimentation with GenAI usage and official organizational coordination in collaborative work settings.
    \item Second, we identify specific challenges that hinder GenAI coordination in collaborative newswork. We demonstrate how the absence of teamwork norms, coordination mechanisms, and shared practices impedes the coordination of GenAI into collaborative newswork.
    \item Third, we outline design and policy implications to inform the development of GenAI systems and organizational practices, fostering more effective coordination of GenAI into collaborative work.
\end{itemize}

\section{Related Work}
We firstly review research on GenAI in the news industry (\autoref{AI}). We then turn to broader studies of collaborations in the newsroom (\autoref{collaboration}). Then, we examine work on GenAI for teamwork and organizational collaboration (\autoref{teamwork}). We conclude by discussing why individual GenAI use cannot be analytically isolated from organizational dynamics and introduce prior HCI and CSCW work on the organizational adoption of emerging technologies (\autoref{orgtech}).

\subsection{GenAI in the News Industry} \label{AI}
This subsection reviews prior work on GenAI in journalism, highlighting both its opportunities and risks (\autoref{opp}) as well as the value tensions that potentially shape its cautious adoption in collaborative newswork (\autoref{value}).

\subsubsection{Opportunities and Concerns of GenAI in Journalism} \label{opp}
Many news organizations now develop GenAI for drafting stories, summarizing reports, suggesting headlines, or even generating interview questions \cite{lewis2025generative,ogola2024between}. Editing tasks are also supported through AI tools that propose alternative phrasings, check stylistic consistency, or flag potential bias \cite{noh2025biassist,wang2025media}. A 2023 global survey found that over 75\% of news organizations across 46 countries were already using AI \cite{journalismai2023survey}. Some leading newsrooms such as China Media Group and Bloomberg are collaborating with AI technologists and experimenting with building tailored GenAI systems, seeking to customize GenAI for journalistic needs \cite{xiao2025might}.

Despite these affordances\add{—functional possibilities that GenAI makes available for journalistic work—}concerns persist. Chief among these are the risks of factual inaccuracy, the amplification of existing biases, and the homogenization of content, where widespread reliance on the same tools may diminish the distinctive voice and editorial identity of individual outlets \cite{siitonen2024mapping,lewis2025generative,thomson2024generative,matich2025old}. Journalists have also voiced existential anxieties about de-skilling and professional displacement, fearing that GenAI may automate precisely those creative and investigative tasks that define the integrity and social value of their work ~\cite{woodruff2024how,siitonen2024mapping}. 

HCI researchers have responded by exploring how GenAI tools can be aligned with newsroom practices \cite{aitamurto2019hci,tseng2025ownership,xiao2025might,nishal2025values,komatsu2020ai}. For instance, Tseng et al. ~\cite{tseng2025ownership} proposed a journalist-centered design space for GenAI, arguing that systems could adapt to the editorial logics and workflows of the profession, rather than impose model-driven efficiencies. Hoque et al. \cite{hoque2024towards} designed a question-answering chatbot to assist journalists in presenting news content to readers in a balanced and objective manner. 

\subsubsection{Tensions Between GenAI and Newswork} \label{value}
Several studies have highlighted how journalistic collective values such as independence, transparency, fairness, and credibility function not just as abstract ideals, but as design requirements for GenAI systems in newsrooms and sources of sociotechnical friction \cite{nishal2025values,moller2024designing,komatsu2020ai,xiao2025might}. Nishal and Diakopoulos \cite{nishal2025values} argue that journalistic values could be treated as central constraints in system development. Other scholars emphasizes that editorial judgment is a situated professional norm that could be maintained even as automation expands in newswork routines \cite{xiao2025might,moller2024designing}. Many newsrooms also formalized their AI policies \cite{hofeditz2025ethical,viner_bateson_2023}. Major organizations such as The Guardian have published explicit editorial guidelines on GenAI, focusing on the core values of the newswork when using GenAI \cite{viner_bateson_2023}. 

Recent studies show that journalists themselves are also experimenting with GenAI in informal, bottom-up ways \cite{wu2024journalists}. Wu \cite{wu2024journalists} characterizes this as a “value-motivated use” framework, where journalists selectively adopt AI only insofar as it supports core editorial standards. In practice, individual journalists used GenAI for tasks such as designing interview questions, assisting with writing, clarifying terminology, summarizing key points, and checking facts. Yet these uses were marked by caution, as journalists had to constantly guard against potential errors, such as hallucinations, that could compromise their work \cite{wu2024journalists}. However, Wu \cite{wu2024journalists} and other journalism studies do not examine how these individual practices relate to broader collaborative norms. 

\subsection{Collaboration in the Newsrooms} \label{collaboration}
Newsrooms are fundamentally collaborative workplaces, where the production of news relies on distributed roles and interdependent tasks. Early ethnographic studies have emphasized that news is not authored by individual journalists alone but is the result of collective coordination \cite{tuchman1973making,cottle2007ethnography,robinson2011journalism}. Collaboration structures how news stories are pitched, developed, verified, and ultimately published \cite{tuchman1973making}.

One core form of collaboration is the front-line journalist–editor relationship. Front-line journalists generate ideas, gather information, and draft stories, while editors provide oversight, frame narratives, and ensure adherence to professional values such as accuracy, fairness, and credibility \cite{frost2010reporting,kolodzy2006convergence,duffy2021out}. Another dimension of collaboration occurs among front-line journalists themselves. Carlson \cite{carlson2009dueling} points out that front-line journalists, even when affiliated with different news organizations, often end up working in shared spaces such as press conferences, media briefings, or public events \cite{dailey2005convergence,carlson2009dueling}. The collaborations of front-line journalists create conditions where information is rapidly exchanged and professional norms are reinforced through peer interactions \cite{crouse2003boys}. 

Collaboration in newsrooms increasingly spans both internal professional boundaries and external organizational ones. Internally, front-line journalists and editors work closely with photojournalists, data journalists, designers, and technical staff to produce multimedia stories, navigating tensions between journalistic norms and technical or esthetic priorities \cite{xiao2025might,wu2021immersive,pavlik2019journalism}. For instance, Xiao et al. \cite{xiao2025might} documented how in-house AI engineers and journalists collaborated to build GenAI tools: engineers providing technical infrastructure, and journalists contributing normative and contextual expertise. Externally, some today's journalists also need to negotiate collaborations with technology companies that supply GenAI systems \cite{dodds2025knowledge,xiao2025let,simon2024escape}; such partnerships then raise new questions about editorial independence \cite{simon2024escape}.

\subsection{GenAI for Teamwork and Organizational Collaboration} \label{teamwork}

Before the rise of GenAI, a long line of research in HCI, CSCW and organizational studies examined how collaborative technologies, from email, shared document platforms, and groupware to chat tools, reshape workplace coordination and communication \cite{rama2006survey,rodden1991survey,kong2024supporting,carasik1988case,wang2022group}. These tools were primarily designed to improve information flow, enable distributed collaboration, and support decision-making across teams \cite{cherbonnier2025collaborative,wang2022group,handel2002chat}. Early AI systems in organizations, such as expert systems or decision-support tools, were typically domain-specific and structured around narrow tasks \cite{billi2025ai,el1993expert,abdolmohammadi1987decision,phillips2012ai}. While useful, these systems rarely participated in open-ended dialogue or generative tasks. In contrast, GenAI introduces a new paradigm: tools that can autonomously generate, reframe, and negotiate meanings. This shift not only expands the scope of automation but also raises new questions about agency and the evolving role of human-AI collaboration \cite{leong2024dittos,shao2025future,IBM_GenAI_2024}.

According to Gartner’s 2024 survey \cite{gartner2024survey}, GenAI has rapidly become the most widely adopted AI category in organizations. GenAI has opened up new opportunities for teamwork and organizational collaborations. At a basic level, GenAI is already widely used to support collaborative functions such as summarizing meeting notes, drafting shared documents, generating ideas, or providing quick answers and even feedback during group discussions. These functions help teams save time, reduce repetitive work, and establish a shared informational baseline \cite{palani2024evolving,han2024teams,takaffoli2024generative,stray2025human, 10.1145/3613904.3642114, doi:10.1056/AIoa2400196}. More recently, however, research has pointed to emerging developments that extend GenAI’s role from a support tool to an active collaborator. For example, Leong \cite{leong2024dittos} introduced the concept of Dittos, AI agents designed as personalized stand-ins that visually and aurally resemble humans and embody their knowledge. These clone agents were found to hold significant potential in team meetings, where they could evoke trust and a sense of presence while providing relevant information to support collective decision-making \cite{johnson2025exploring,hu2025your}.

\subsection{Emerging Technologies, Organizational Adoption, and the Limits of Individual Use} \label{orgtech}

\add{A long tradition in CSCW and HCI shows that emerging technologies do not remain individual for long. Although new tools often enter workplaces through personal experimentation, these practices invariably spill over into collaborative norms \cite{grudin1988cscw,beane2019shadow,cha2025understanding}. Technologies such as email, instant messaging, shared document platforms began as “personal tools,” yet their informal uptake had organizational consequences well before formal policies were introduced \cite{wallace2004internet,anders2016team,bhat2021infrastructuring}.}

\add{This scholarship demonstrates that individual use is never merely individual in collaborative environments. When collaborators rely on different tools, use them unevenly, or conceal their usage altogether, articulation work increases and invisible repair labor emerges \cite{jarrahi2015theorizing,grudin1988cscw}. Even non-use plays an active role: strategic refusal, avoidance, and conditional use express professional values and risk perceptions, which then shape collective workflows and expectations \cite{selwyn2005whose,garg2019you,cha2025understanding,baumer2013limiting}. As prior studies show, these heterogeneous stances toward a technology reshape teamwork long before an organization formally acknowledges or regulates the tool \cite{baba1999dangerous,10.1145/3613904.3642114,li2025vibe}. Research on organizational adoption and governance of emerging technologies further demonstrates that when bottom-up use becomes widespread, organizations could grapple with new questions around visibility, accountability, workflow coordination, and legitimacy \cite{khovanskaya2020bottom,banks2016team,karunakaran2024frontline}. }

\add{Against this backdrop, GenAI could not be treated as an exception. Like earlier waves of workplace technologies, GenAI initially appears as an individualized, private, or experimental practice \cite{wu2024journalists}. Yet because journalism is a deeply interdependent domain, where editorial quality, fact-checking, and organizational reputation rely on coordinated judgment, individual GenAI use inevitably produces collective effects. Hidden or uneven GenAI use alters how stories are drafted, how decisions are made, and how outputs are evaluated. It also amplifies well-documented challenges around visibility, provenance, and accountability \cite{wu2024journalists}.}

\add{Therefore, this study follows directly from decades of HCI and CSCW research demonstrating that once technologies become entangled with collaborative work, questions about governance, transparency, and shared practices necessarily arise. Integrating GenAI into this tradition allows us to see that the core issue is not whether individual journalists use GenAI, but how these individual practices reshape organizational norms and create new forms of interdependence that require collective management.}

\section{Background: GenAI in the Chinese News Industry}
In China, the news industry represents a distinctive setting due to its combination of rapid technological experimentation and strong regulation. The Chinese government has encouraged newsrooms to experiment with GenAI, including initiatives such as developing in-house GenAI systems through continuous recruitment of AI specialists and collaborations with technology companies and university labs \cite{xiao2025let,chinaxinwen2024aimedia}, reflecting a broader ambition to position China at the forefront of AI-driven media transformation. \add{At the same time, even as organizations pursue large-scale GenAI system development \cite{xiao2025let,chinaxinwen2024aimedia}, individual journalists have begun adopting GenAI tools in their daily practice \cite{liu2025culture}, creating a parallel bottom-up trajectory of experimentation that coexists with top-down infrastructural ambitions.}

The Chinese news industry has made notable progress in the design of the GenAI system for news work.\footnote{Yet, it is worth noting that the development of these newsroom-specific GenAI systems has not been particularly transparent. For example, most Chinese newsrooms have not disclosed whether their “proprietary” models are entirely self-developed or fine-tuned from existing commercial or open-source foundations. All descriptions summarized here in the background section are directly from the state news reports about Chinese GenAI development in newsrooms, which limits independent verification of the underlying technical processes in this paper.} In July 2023, Xinhua News Agency \cite{chinaxinwen2024aimedia} released its proprietary large language model. The model was soon applied in coverage of the Chengdu Universiade, where Xinhua combined GenAI-based image and text generation to create and publish AIGC news videos \cite{chinaxinwen2024aimedia}. Around the same time, China Media Group (CMG) introduced the “CMG Media GPT,” the first large-scale AI model in China dedicated specifically to audiovisual content production \cite{chinaxinwen2024aimedia,lang2024qianqiu_shisong}. Even regional newspapers also accelerated their investment in domain-specific GenAI models. Zhejiang Radio and Television Group launched an AIGC-based platform to support professional news production and used it to publish news documentaries \cite{chinaxinwen2024aimedia}. 

These developments have drawn considerable attention in China’s media policy discourse. As noted by the China Press, Publication, Radio, Film and Television Journal \cite{chinaxinwen2024aimedia}, which reflects the government’s stance and is directly overseen by the State Administration, such GenAI experiments highlight the need for journalists to develop new paradigms of human–AI collaboration in their daily newswork. Meanwhile, regulatory oversight in China has intensified. In 2023 the authorities released new rules governing the use of AI in news, which combined stricter accountability with incentives for news organizations to expand the use of GenAI in content production \cite{cac2023}. Starting from September 1, 2025, the Chinese government requires that all AI-generated content be clearly labeled. Any organization or individual found to maliciously delete, alter, forge, or conceal such labels will be considered in violation of the regulation and subject to penalties \cite{cctv2025aicontent}. 

\section{Method}

\add{Our study investigates how individually adopted GenAI practices begin to intersect with team coordination and collaborative newswork, even when organization-level coordination remains in its early stages.} We conducted 27 in-depth interviews with senior newsroom managers, editors, and front-line journalists in China. All participants were affiliated with news organizations that had initiated early-stage projects involving the development or deployment of GenAI tools. Each journalist had direct experience with GenAI in their collaborative newswork. In the following, we outline our participant recruitment strategy, the design and conduct of the interviews, and our approach to data analysis.

\subsection{Participants}

We used purposive sampling \cite{firdaus2024small} to recruit journalists from potential Chinese newsrooms actively engaged in early-stage GenAI experimentation or implementation. We ultimately interviewed 27 news practitioners (see \autoref{tab:Participant}). All participants were working in news organizations that had either (a) integrated general-purpose large language models into their workflows, (b) developed domain-specific GenAI models tailored for journalism (e.g., models fine-tuned on news language, ethics, and practices to support journalism broadly), or (c) implemented newsroom-specific GenAI tools designed for concrete editorial tasks such as article drafting, summarization, fact-checking, or text and video news production. A prerequisite for inclusion was direct experience with GenAI in daily newswork. 

\begin{table*}[ht]
  \centering
  \caption{Participant Demographics}
  \label{tab:Participant}
  \setlength{\tabcolsep}{6pt}
  \begin{tabularx}{\textwidth}{ccccccc}
    \toprule
    \textbf{ID} & \textbf{Gender} & \textbf{Age} &\textbf{ Journalism Experience} & \textbf{Position} & \textbf{GenAI Experience} &\textbf{ GenAI Use Frequency} \\
    \midrule
    P1  & Female & 48 & 22 years & Editor/Senior Manager & 1, 2 & Several times per day \\
    P2  & Female & 42 & 19 years & Editor/Senior Manager & 1 & Several times per week \\
    P3  & Male   & 45 & 23 years & Editor/Senior Manager & 1 & Several times per week \\
    P4  & Female & 38 & 13 years & Editor & 1 & Several times per week \\
    P5  & Female & 34 & 11 years & Editor & 1, 2 & Several times per day \\
    P6  & Female & 33 & 10 years & Editor & 2 & Several times per week \\
    P7  & Male   & 37 & 10 years & Editor & 1, 2 & Several times per day \\
    P8  & Male   & 35 & 12 years & Editor & 2 & Several times per week \\
    P9  & Male   & 29 & 3 years  & Editor & 2 & Several times per week \\
    P10 & Male   & 28 & 3 years  & Front-line Journalist & 2 & Several times per day \\
    P11 & Female & 26 & 2 years  & Front-line Journalist & 2 & Several times per week \\
    P12 & Female & 24 & 2 years  & Front-line Journalist & 2 & Several times per week \\
    P13 & Male   & 32 & 8 years  & Editor/Front-line Journalist & 2 & Several times per day \\
    P14 & Female & 30 & 6 years  & Editor/Front-line Journalist & 2 & Several times per day \\
    P15 & Female & 28 & 3 years  & Front-line Journalist & 1, 2 & Several times per day \\
    P16 & Female & 26 & 2 years  & Front-line Journalist & 2 & Several times per day \\
    P17 & Male   & 25 & 1 year   & Front-line Journalist & 2 & Several times per week \\
    P18 & Male   & 28 & 3 years  & Editor/Front-line Journalist & 1, 2 & Several times per day \\
    P19 & Male   & 25 & 1 year   & Front-line Journalist & 2 & Several times per day \\
    P20 & Male   & 25 & 1 year   & Front-line Journalist & 2 & Several times per day \\
    P21 & Male   & 25 & 1 year   & Front-line Journalist & 2 & Several times per week \\
    P22 & Male   & 25 & 2 years  & Front-line Journalist & 2 & Several times per day \\
    P23 & Male   & 26 & 1 year   & Front-line Journalist & 2 & Several times per day \\
    P24 & Female & 26 & 2 years  & Front-line Journalist & 2 & Several times per week \\
    P25 & Female & 26 & 1 year   & Front-line Journalist & 2 & Several times per day \\
    P26 & Female & 25 & 1 year   & Front-line Journalist & 2 & Several times per day \\
    P27 & Male   & 25 & 1 year   & Front-line Journalist & 2 & Several times per day \\
    \bottomrule
  \end{tabularx}
  \vspace{1ex}
  \begin{minipage}{\textwidth}
    \footnotesize
    \textbf{Note:}  
    GenAI experience codes:  
    1 = Participation in organizational GenAI strategy or development planning;  
    2 = Use of GenAI tools in daily newswork.   
  \end{minipage}
\end{table*}

To enrich our data with diverse perspectives and levels of organizational embedding, we sampled across different seniority levels, including editors, senior managers, and front-line journalists. Given our research goal of understanding how GenAI could be integrated into collaborative newswork, we purposely recruited participants who had already engaged with GenAI in their professional roles. Rather than studying those unfamiliar with GenAI and never thinking about future collaborative newswork around GenAI, we sought insight from those already reflecting on its collaborative implications. 17 participants were recruited through the authors’ personal networks, drawing on prior experience in journalism in China, while the remaining 10 were recruited through snowball sampling based on participants’ professional contacts. These participants were not only early adopters of GenAI, but also embedded in organizational environments at the forefront of GenAI innovation. 

Participants' journalism experience ranged from one to over twenty years, and they represented a spectrum of positions including front-line journalists (n=16), editors (n=8), and senior managers in newsrooms (n=3). Some participants simultaneously held multiple roles within their organizations, as shown in \autoref{tab:Participant}. Front-line journalists offered grounded insights into how GenAI tools affect day-to-day reporting in collaborations. Mid-level editors, positioned between strategic directives and ground-level implementation, shared perspectives on coordinating GenAI  into team workflows and balancing technological adoption with editorial standards. Senior managers contributed a broader organizational view, describing strategic rationales behind GenAI coordination, concerns about institutional credibility, and long-term visions for human–AI collaboration in journalism. Some editors also used GenAI to support their editorial tasks, and some front-line journalists occasionally participated in organizational discussions around GenAI strategy.

\add{All participants were drawn from regional news organizations in China, where digital news production has become a central component of daily editorial work. Their newsrooms operate in a dual landscape: on one hand, journalists engage in bottom-up, individualized experimentation with GenAI; on the other hand, organizations are investing heavily in top-down development of proprietary AI tools. Yet despite these parallel trajectories, no formal mechanisms currently exist to coordinate individual GenAI use within collaborative team workflows.}

\add{Across these newsrooms, participants regularly engaged in digitally mediated reporting, editing, multimedia production, and platform content distribution. Their primary collaborators included editors responsible for narrative framing and quality assurance, fellow front-line journalists who jointly covered events or exchanged source materials, and, in some organizations, in-house AI engineers and product teams who developed or maintained emerging GenAI tools. Although specific collaboration structures varied by newsroom size and workflow arrangements, most participants worked within small reporting teams or editor–journalist pairs, coordinated through shared content management systems, group chats, and daily editorial meetings. In several cases, participants also had overlapping working relationships, for example, junior journalists who collaborated with multiple editors, or editors who simultaneously supervised and co-produced content with different reporting teams.}

\add{Participation in the study was entirely voluntary, and no financial or material compensation was provided, as participants chose to participate based on their genuine interest in GenAI’s implications for journalism rather than external rewards.} To protect participant confidentiality, all individuals were anonymized and assigned pseudonyms in the form of participant IDs ranging from P1 to P27. 

\subsection{Study Design}
To explore how journalists experience, negotiate, and reflect on the coordination of GenAI into their teamwork, we conducted 27 semi-structured interviews. Each interview lasted approximately 90–120 minutes and was conducted remotely. \add{A complete interview protocol, including phase-by-phase guiding questions, is provided in the \autoref{protocol} for reference.}

Each interview was structured into three phases. First, participants were asked to recall and describe concrete experiences with GenAI in their newsroom work, with particular attention to how these experiences unfolded in collaborative contexts. Second, the interview shifted to expectations. We asked participants to imagine what ideal GenAI coordination would look like in collaborative newswork. Example questions included: \textit{“In your ideal scenario, how would GenAI assist not only you but also your team?”} and \textit{“What would make a GenAI tool feel reliable in teamwork?”} Finally, participants were invited to reflect critically on the gap between their collaborative ideals and current realities. We asked them to identify what prevents GenAI from being coordinated into team workflows, and what organizational or technical changes might better support GenAI in collaborative newswork. 

\subsection{Data Analysis}

\add{We analyzed the interviews using reflexive thematic analysis \cite{braun2019reflecting}, following an iterative, interpretive, and team-based approach. Our analytic process unfolded across three stages.}

\add{\textit{Stage 1: Immersion and initial coding.} The researchers read the transcripts in full to build familiarity with the dataset. Three researchers independently conducted open coding on a shared subset of transcripts, noting instances of GenAI use, collaboration challenges, informal workarounds, value tensions, and moments of uncertainty or ambiguity. These codes were intentionally descriptive and numerous, capturing as much detail as possible without imposing early structure. The team then met to compare initial codes, discuss divergent interpretations, and collaboratively construct a preliminary coding frame that reflected our emerging analytic interests.}

\add{\textit{Stage 2: Systematic coding and iterative refinement.} Using the initial coding frame, the full dataset was coded by the research team. Following the principles of reflexive qualitative analysis, disagreement was treated not as a reliability problem but as a productive site for interpretation \cite{morgan2020iterative}. Researchers wrote analytic memos to trace emerging patterns, theoretical questions, and surprising contradictions in participants’ accounts. We met regularly to refine code definitions, merge overlapping codes, elevate promising ones, and prune those that did not hold analytically. These discussions helped us critically examine our assumptions and ensure that our interpretations remained grounded in, rather than imposed onto, the data.}

\begin{figure*}[t]
    \centering
    \includegraphics[width=\linewidth]{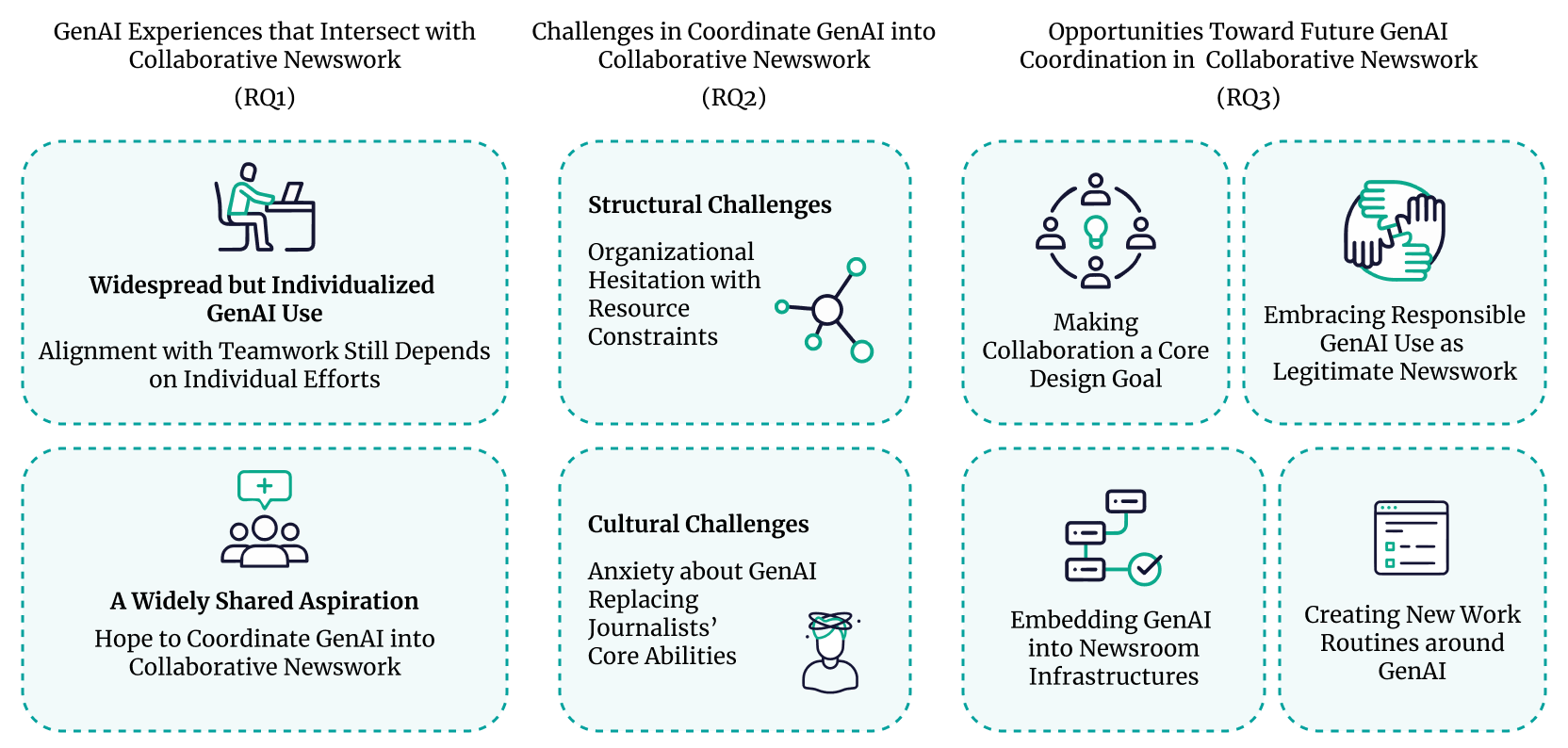}
    \caption{Overview of the Findings Structure}
    \label{fig:findings-overview}
\end{figure*}

\add{\textit{Stage 3: Theme construction and validation.} Once coding was complete, we clustered related codes into broader themes that captured recurrent patterns across participants’ experiences. In line with reflexive thematic analysis, themes were understood as meaning-based patterns, not frequency counts. We repeatedly returned to raw excerpts to validate the coherence and distinctiveness of each theme, testing whether they adequately represented participants’ experiences and reflected the tensions between individual GenAI experimentation and collaborative newsroom workflows. The final themes, detailed in \autoref{routine}, \autoref{chall}, and \autoref{opportunity}, are also summarized in Figure~\ref{fig:findings-overview}.}

\subsection{Positionality Statement}

\add{We acknowledge that our backgrounds, methodological training, and prior professional experiences shape how we approached this study and interpreted participants’ accounts. Our research team is trained in Human–Computer Interaction, Communication, Journalism, and Organizational Studies, and several team members have prior work experience in digital journalism and human–AI collaboration. Two researchers previously worked in or closely studied Chinese newsrooms, which gives us familiarity with the professional norms, political sensitivities, and hierarchical structures that characterize this environment.}

\add{Throughout the study, we engaged in deliberate reflexive practice, following approaches in HCI and CSCW that encourage researchers to critically examine their assumptions throughout the research process \cite{kuo2023understanding}. We continuously asked: (1) How might our own expectations about GenAI influence the questions we asked or the patterns we foregrounded? (2) What biases do we bring from our familiarity with newsroom culture or from our broader research agendas on human–AI collaboration? (3) Who benefits from this research, and whose perspectives might remain underrepresented? Each team member documented these reflections during analysis, and we compared them during coding meetings. These reflexive discussions made us more attentive to the multiple interpretations of journalists’ practices—for example, recognizing that cautious GenAI use could stem not only from technical distrust but also from hierarchical accountability, political sensitivity, or cultural norms around authorship. We aimed to conduct our research with humility, respect for participants’ expertise, and an awareness of how our positionality shaped the knowledge we produced.}

\section{Findings: Still Individual Use, but with Expectations for Coordinating GenAI into Collaboration (RQ1)} \label{routine}

Our findings reveal that although GenAI use is already widespread at the individual level in our sampled newsrooms, it is highly personalized, informal, and largely invisible to colleagues (\autoref{wide}). Yet, these individualized practices have not yet coalesced into teamwork or newsroom-wide standards for future collaborations. Even so, many journalists described actively working to ensure that their own GenAI use, and that of their teammates, remained aligned with professional values in collaborative work by individual efforts. Despite the lack of formal coordination, journalists nonetheless express a strong desire for GenAI to become a part of their collaborative newswork  (\autoref{motivations}).

\subsection{Widespread but Individual Use of GenAI in Chinese Newsrooms} \label{wide}

Across our interviews, participants expressed a wide spectrum of attitudes toward the use of GenAI in the newsroom, ranging from pessimistic to optimistic. Some viewed GenAI as a potential threat to journalistic rigor, fearing that it might \textit{“destroy the entire editorial process, make journalists lazier and less reflective, and ultimately lead them to rely on GenAI-generated text at the expense of their own writing skills”} (\( P22_{\text{male, journalist}} \)). Others held a more hopeful outlook, believing that \textit{“GenAI can significantly improve journalistic productivity by saving time on tedious tasks and freeing up energy for meaningful investigative work”} (\( P25_{\text{female, journalist}} \)). 

Despite these differences, there was broad consensus on one point: the use of GenAI by individual journalists is already widespread in 2025 in China and likely irreversible. \( P2_{\text{female, editor/manager}} \), a senior manager in her newsroom who was responsible for major organizational decisions and frequent communication with government authorities, cited a 2024 report by the research team from Xinhua News Agency \cite{xinhua2024responsibility}, the China’s most prominent state newspaper, stating that \textit{“very few newsrooms have expressed total rejection of GenAI in organizational level. Most are simply waiting for the right moment, for the GenAI technology to mature enough to be formally introduced into collaborative newswork.”} 

\add{\( P15_{\text{female, journalist}} \) offered a complementary perspective grounded in her everyday work. \textit{“As far as I know, many of my colleagues are already using GenAI. We use it to brainstorm and to revise our early drafts. For me, the final article is nearly no similarity compared to the AI draft because I revise it so much, but GenAI is essential in the very beginning, when getting the first version down is the hardest part for me.”} In her view, when journalists talk about the ‘widespread’ use of GenAI, this does not mean that everyone uses GenAI all the time, nor that GenAI is involved in every story. Rather, \textit{“people inevitably find themselves turning to GenAI at certain moments,”} as she described.}

\add{Journalists’ engagement with GenAI was selective, situational, and guided heavily by personal judgment. Some (e.g., \( P19_{\text{male, journalist}} \), \( P20_{\text{male, journalist}} \)) relied on GenAI multiple times a day for drafting or summarizing; others used it only when pressed for time (e.g., \( P17_{\text{male, journalist}} \)), and a few avoided it entirely for tasks they viewed as requiring professional integrity or narrative control (e.g., \( P21_{\text{male, journalist}} \)). Among our participants, they commonly felt, most journalists draw on GenAI when they feel it is needed, yet such on-demand use is still widespread across individuals.} 

\add{Yet it is precisely this fragmented and uneven individualized use that emerged as consequential in our study. In a highly interdependent newsroom environment, the core challenge was the pervasive uncertainty it created: journalists could rarely tell whether, when, or how their colleagues had used GenAI. This ambiguity set off what many participants described as a subtle chain of mutual suspicion. As \( P7_{\text{male, editor}} \) put it, \textit{“Sometimes I try to edit a draft and I really wonder: did they use GenAI here or not? I have no way to know, but I have to check it as if they did.”} Journalists were unsure whether drafts they received from peers contained undisclosed AI-generated passages; editors often wondered whether a stylistic shift or factual slip reflected GenAI involvement; and teammates speculated about how much additional verification they were expected to perform. In the \autoref{motivations}, we show how these uncertainties and suspicion chains motivated journalists to imagine ways of bringing individualized GenAI practices into more coordinated and transparent forms of collaborative and organizational coordination.}

It is worth noting that in our study, all direct interactions between journalists and GenAI took the form of conversational exchanges with chat-based GenAI systems such as Qwen or DeepSeek. Participants typically engaged GenAI by typing prompts in natural language, requesting draft texts, summaries, translations, or stylistic adjustments. In other words, GenAI was used as an interactive writing partner rather than through more technical means such as API calls, system integration, or backend customization. This use of GenAI was integrated directly into the writing and editing, while also reinforcing its status as a largely private, individualized practice that can be easily concealed from colleagues and editors.

\add{It is also important to note that, despite the rapid expansion of multimodal GenAI capabilities, participants overwhelmingly described using GenAI primarily for text-based tasks rather than image or video generation. This pattern was not due to a lack of technical availability, as many journalists were familiar with GenAI image and video tools but reflected deeper professional and regulatory constraints. In Chinese newsrooms, AI-generated images or videos are largely considered inappropriate for routine news production and are only acceptable when intentionally staged as AIGC content or published within clearly experimental formats (e.g., special AI zones). Outside of these contexts, AI-generated visuals are widely viewed as carrying political, ethical, and credibility risks: they can be mistaken for manipulated media, trigger public suspicion, or violate strict editorial norms around visual authenticity. As several participants noted, \textit{“AI pictures are a taboo in everyday news, we are not allowed to use that in our work without permission”} (\( P11_{\text{female, journalist}} \)) unless their artificial nature is the point of the story. By contrast, text generation was seen as easier to regulate, revise, and ‘humanize’ through subsequent editing, making it the dominant modality through which GenAI entered everyday newswork.}

\subsection{Expectation for Better Coordinating GenAI into Collaborative Newswork} \label{motivations}

\add{Across the newsrooms we studied, journalists firstly described a shared sense that GenAI could no longer purely remain an individualized, private tool because this arrangement placed disproportionate burdens on individual journalists and editors.} In our sampled newsrooms, maintaining the alignment between GenAI usage and journalistic expectations currently depends almost entirely on individual effort. Generally, front-line journalists typically draft stories, exchange leads with peers, and submit copy to editors, who then review, revise, and make publication decisions. Within this layered process, journalists experimenting with GenAI could not only ensure the appropriateness of their own outputs but also anticipate how these outputs will intersect with editors’ judgments and colleagues’ contributions.

\add{Most participants described the extra effort this entailed: double-checking GenAI-generated text before passing it along, quietly monitoring whether teammates might have relied on GenAI, and compensating for possible errors to avoid undermining collective products. Editors, too, were not merely gatekeepers reacting to GenAI-produced content; some such as \( P5_{\text{female, editor}} \) and \( P7_{\text{male, editor}} \) also adopted GenAI tools to support their own editorial work. \add{In both roles, however, using GenAI created a dual layer of responsibility: journalists and editors had to remain accountable for the AI-assisted text they produced themselves, while also devoting additional effort to scrutinizing the materials they received from colleagues. According to \( P7_{\text{male, editor}} \), \textit{“this scrutiny now extended to stylistic and tonal indicators, especially given the recent rise of public vigilance toward AI-generated writing”.} \( P7_{\text{male, editor}} \) noted, \textit{“even minor shifts in phrasing could draw audience suspicion, placing heightened pressure on editors, who hold final responsibility of any AI-related risks before publication.”}}}

\add{Participants further emphasized that GenAI’s largely private and disconnected use created pervasive uncertainty in collaborative work, reinforcing their desire for more coordinated practices.} Few had insight into how others were deploying GenAI in daily work, and most described their usage as \textit{“private” }(e.g., \( P9_{\text{male, editor}} \), \( P13_{\text{male, editor/journalist}} \), \( P14_{\text{female, editor/journalist}} \)) or \textit{“informal”} (e.g., \( P5_{\text{female, editor}} \), \( P7_{\text{male, editor}} \), \( P16_{\text{female, editor}} \)), a backstage tool rather than something openly discussed within teams. This produced hidden coordination labor: journalists and editors worried about how much extra verification was needed, and whether colleagues had unknowingly relied on GenAI in ways that could introduce subtle errors.

\add{Moreover, because these uncertainties threatened not just individual journalists but team- and organizational-level outcomes, participants argued that GenAI required collective guardrails. Several participants (n=11), described deep concerns that the lack of shared standards allowed GenAI to introduce errors that would impact collective news products and organizational credibility. Front-line journalists worried that undisclosed AI use could lead to \textit{“mistakes being traced back”} to them, exposing them to \textit{“individual blame”} and \textit{“disciplinary scrutiny”} (\( P14_{\text{female, editor/journalist}} \)); editors feared reputational damage because they are \textit{“ultimately accountable for publication quality”} (\( P8_{\text{male, editor}} \)); managers were concerned about political, regulatory, and ethical consequences if AI-assisted content contaminated organizational outputs, \textit{“especially in China that are so sensitive, even small mistakes can become dangerous"} (\( P1_{\text{female, editor/manager}} \)).}

\add{These perceived risks also highlighted a broader issue of fairness, motivating participants to call for more explicit organizational practices that would distribute responsibility more equitably. Many felt that the current ambiguity enabled organizations to\textit{ “offload risks onto individuals”} (\( P12_{\text{female, journalist}} \)), especially junior journalists and editors, while avoiding institutional accountability. }

\add{Finally, participants hoped that bringing GenAI into collective newswork would foster transparency, peer learning, and shared norms needed for responsible use.} Many (e.g., \( P3_{\text{male, editor/manager}} \), \( P12_{\text{female, journalist}} \), \( P20_{\text{male, journalist}} \)) called for open conversations about GenAI’s future role in newswork, rather than allowing it to remain in a gray zone of private experimentation. They envisioned shared discussions about collaborative best practices, institutional expectations, and ethical tradeoffs as foundational for building trust and making GenAI a legitimate part of teamwork. 

In short, participants’ motivations for coordinatingg GenAI into collaborative newswork stemmed from a combination of individual burdens, hidden coordination work, perceived organizational risks, concerns about fairness, and the need for shared norms.

\section{Findings: Challenges in Coordinating GenAI into Collaborative Newswork (RQ2)} \label{chall}
In the \autoref{routine}, we demonstrated that the current use of GenAI in newsrooms is highly individualized. Journalists engage with GenAI tools in personal, often private ways, with limited visibility into how their colleagues are using similar technologies. This fragmented usage stands in sharp contrast to the collaborative nature of journalistic work, which typically relies on shared norms, editorial review, and team-based production. In this section, we investigate \add{why current Chinese newsrooms not yet developed mechanisms to coordinate}. It is important to clarify that our findings focus only on traditional newswork and do not extend to journalist-technical companies or journalists-AI engineers partnerships.

These challenges in coordinating GenAI into collaborative newswork can be broadly grouped into two interrelated domains: structural and cultural. Structurally, many newsrooms lack formal coordination mechanisms for GenAI usage, making it difficult to establish shared visibility or accountability across team members (\autoref{stru}). Culturally, journalists often regard their GenAI use as private, experimental, or even stigmatized, leading to reluctance in disclosing their practices or outputs to peers and supervisors (\autoref{Culture}). 

\subsection{Structural Challenges: Lack of Organizational Mechanisms for Integration} \label{stru}
While individual journalists have actively experimented with GenAI tools, efforts to embed these practices into newsroom-level collaborative norms face critical structural barriers. These structural challenges stem from a lack of organizational incentives or mandates to coordinate GenAI usage (\autoref{first}), insufficient technical infrastructure to support collaborative workflows (\autoref{second}), and heightened public scrutiny that discourages formal coordination without robust safeguards (\autoref{third}).

\subsubsection{First, news organizations often lack structural incentives or mandates to coordinate GenAI usage in teamwork.} \label{first}

\add{In the context of China’s news industry, organizational priorities are often oriented toward meeting political expectations, demonstrating technological capability, and maintaining institutional legitimacy. Within this configuration, GenAI is primarily evaluated as a strategic or symbolic asset, used to signal innovation, efficiency, and alignment with policy agendas, rather than as a technology requiring sustained coordination within collaborative newswork. As a result, questions of how GenAI should be collectively governed, integrated into team-based workflows, or aligned with shared editorial norms have received comparatively little attention. Participants noted that this emphasis on institutional signaling over collaborative coordination leaves front-line teams to navigate GenAI-related risks individually, without clear, collectively articulated guardrails.}

Against this backdrop, although many newsroom leaders (e.g., \( P1_{\text{female, editor/manager}} \), \( P2_{\text{female, editor/manager}} \), \( P3_{\text{male, editor/manager}} \)) are aware that individual journalists within their organizations are already experimenting with GenAI tools, they rarely translate this awareness into concrete action at the team level. \textit{“We’ve done a lot to build internal GenAI systems, which is mainly to satisfy government expectations,” }\( P13_{\text{male, editor/journalist}} \) explained, \textit{“but unfortunately, we don’t yet have a coherent collaboration guide about how to use GenAI for our journalists. No one really knows how it could enhance teamwork.”} From a broader organizational perspective, \( P1_{\text{female, editor/manager}} \), a senior manager, articulated this divide more explicitly. \( P1_{\text{female, editor/manager}} \) emphasized that newsroom-level initiatives, often launched in response to national calls, tend to prioritize the development of specific GenAI systems, rather than paying close attention to how journalists incorporate GenAI into their daily practices.

\add{This symbolic–practical gap also highlights a deeper structural hesitation: many organizations remain uncertain about what it would mean to meaningfully integrate GenAI into journalistic labor.} Several editors (n=7) voiced explicit concerns that formalizing GenAI use might unintentionally signal an endorsement of automation, thereby threatening long-standing norms around authorship, creativity, and editorial authority. \( P13_{\text{male, editor/journalist}} \) expressed this dilemma: \textit{“If we say it’s okay to use GenAI, would that encourage journalists to rely on it too much, instead of their own judgment? And how would our audience perceive this? Maybe they’d think, ‘Oh, this newspaper is just letting AI write the news now,’ or ‘Their journalism could be so poor that AI can step in.’ Some news audiences still carry a lot of negative impressions about GenAI.”} This institutional hesitation creates a structural vacuum: GenAI becomes a tolerated but unregulated practice, operating in a gray zone between tacit acceptance and formal prohibition.

\add{Compounding this uncertainty, even newsrooms that recognize potential collaborative value often lack the expertise, confidence, or conceptual frameworks needed to operationalize GenAI at the teamwork level.} As senior manager \( P3_{\text{male, editor/manager}} \) put it bluntly: \textit{“We don’t even know how to try. What does GenAI collaboration actually mean?”} \( P3_{\text{male, editor/manager}} \) contrasted this ambiguity in collaboration with the current state of highly individualized GenAI use, where accountability remains clear. \textit{“Look, if everyone is using GenAI in a personal and unofficial way, and something goes wrong, we know who to hold responsible: it’s the individual journalist, right?”} However, \( P3_{\text{male, editor/manager}} \) raised concerns about what might happen if the newsroom were to formalize guidelines for how GenAI could be coordinated into team workflows. \textit{“Now imagine we create an official framework for collaboration between human journalists and GenAI. Let’s say we even publish a clear policy about how GenAI shall be incorporated into teamwork. Then if a problem arises, who’s accountable? The journalist might say, ‘I was just following the policy.’ So where does that leave us?”} 

\add{These concerns illustrate a core tension facing many newsrooms. On the one hand, organizations hesitate to define GenAI collaboration for fear of enabling blame deflection, complicating accountability, or introducing institutional risk. On the other hand, leaving GenAI use uncoordinated preserves individual accountability but fragments practices, creating inconsistent, opaque, and difficult-to-scale workflows. As a result, many newsrooms remain trapped in a structural limbo—aware of GenAI’s potential to reshape journalistic collaboration, yet uncertain or unwilling to articulate what responsible, organization-level coordination would actually entail.}

\subsubsection{Second, the technical infrastructure of most Chinese newsrooms poses significant challenges to building collaborative GenAI norms.} \label{second}

Most GenAI tools were accessed individually via external platforms, such as Qwen Chat, Doubao, or DeepSeek. There are no built-in mechanisms in newsrooms to track when, where, or how GenAI has been used in the production of a specific story. As \( P2_{\text{female, editor/manager}} \) explained, \textit{“People are mostly using GenAI on external platforms, which makes it hard for us to trace how they’re using it. We’re trying to build in-house GenAI systems with technical companies, but those are still in pilot stages. In most cases, journalists are still using external GenAI tools.”}

In addition to traceability issues, GenAI tools lack shared interfaces that support collaborative review and editing. Unlike tools such as Google Docs, which facilitate real-time co-editing and inline commenting, most GenAI platforms are designed for single-user interaction. Journalists often copy and paste GenAI outputs into shared documents or use them as prompts for team discussions without revealing the prompts they used, the iterations they tried, or any potential misunderstandings by the model. These limitations are compounded by the fact that GenAI is fundamentally a black-box system. Editors, especially senior editors without a technical background, often lack the AI literacy needed to critically evaluate GenAI outputs. As \( P3_{\text{male, editor/manager}} \) remarked, \textit{“On one hand, we don’t know how the journalist interacted with the GenAI. On the other hand, we don’t know how the GenAI came up with its response. It’s like two black boxes stacked on top of each other, which creates real challenges for editors.” } \add{This “double black box”—the opacity of human interaction layered atop the opacity of model reasoning—captures a unique form, where neither the human process nor the machine process is inspectable, auditable, or mutually accountable.} This dual opacity of both human prompting and machine output undermines editorial confidence and creates a major barrier to coordinate GenAI into collaborative newswork.

The organizational structure of most newsroom tech teams further exacerbates these technical constraints. In our interviews, \( P1_{\text{female, editor/manager}} \), \( P2_{\text{female, editor/manager}} \), and \( P3_{\text{male, editor/manager}} \) all acknowledged that their teams of developers were typically small, under-resourced, and primarily focused on building core GenAI functionalities, such as fine-tuning models or ensuring data compliance, rather than designing interfaces that support multi-user collaboration. As \( P1_{\text{female, editor/manager}} \) explained, \textit{“Right now we’re still at the stage of putting food on the table, not worrying about how good the meal is. We need to get a basic GenAI platform running internally in the newsroom first to meet Chinese government's expectations.”} \( P1_{\text{female, editor/manager}} \) further mentioned, \textit{“Only then can we think about building collaborative platforms. So for now, we let journalists explore freely and use whichever GenAI tools they prefer, instead of developing a system for coordinated use. We just don’t have the capacity.”} 

\subsubsection{Third, as nearly all editor participants in our study emphasized, reviewing GenAI-generated content and anticipating its potential risks has become increasingly difficult, especially as public awareness and sensitivity toward AI-generated news continues to grow.} \label{third}

This enhances the challenges for front-line journalist-editors collaborations around GenAI and the hesitations to promote an official guideline to coordinate GenAI into collaborative newswork. Editors are no longer only concerned with factual accuracy or stylistic quality; they could also anticipate how GenAI usage will be perceived by increasingly critical audiences.

\( P14_{\text{female, editor/journalist}} \), for example, regularly monitors discussions on Chinese social media platforms such as Xiaohongshu and Weibo. She observed that readers are becoming more vigilant and suspicious about the presence of GenAI in journalistic content. As she described, \textit{“In the past, it was enough to ensure that the content met quality standards. Audiences wouldn’t notice or care whether AI was involved. But this year, especially in 2025, that has changed. You’ll find a lot of social media users actively trying to identify whether certain news articles were written by GenAI. And if they think the content was AI-generated, they feel disappointed. Sometimes these posts go viral, with thousands of likes, shares, and comments.”}

This shift in public scrutiny introduces reputational risks for newsrooms and adds further pressure on editorial teams to detect GenAI usage, that are already structurally and technically difficult to manage, as discussed earlier. As a result, newsroom leaders such as \( P1_{\text{female, editor/manager}} \) and \( P2_{\text{female, editor/manager}} \) are particularly cautious about openly coordinating GenAI into editorial newswork. Many fear that any premature or poorly safeguarded disclosure of GenAI use could trigger criticism, undermine credibility, and damage institutional trust. 

\subsection{Cultural Challenges: Rethinking Journalists’ Identities in Collaborative Newswork Around GenAI} \label{Culture}
Culturally, journalists often perceive the use of GenAI as private, experimental, or even stigmatized, leading to reluctance in disclosing their practices or outputs to journalist peers and editorial supervisors. This secrecy poses a major obstacle to coordinating GenAI into collaborative newsroom workflows. Many journalists do not yet regard GenAI as a legitimate component of professional practice, but rather as a gray-area personal assistant. Because these practices are seen as outside the boundaries of “real journalism,” they are hidden from view, preventing the development of shared norms.

\( P24_{\text{female, journalist}} \) explicitly described using GenAI as a form of professional shame, likening it to \textit{“cutting corners when no one is watching.”} She explained that even if the final output is polished through human editing, the mere fact that the first draft was machine-generated made her feel like she had \textit{“cheated on an exam.”} For her, GenAI was not a neutral tool, but a shortcut that undermined the pride she took in crafting language herself. \textit{“It’s like copying someone else’s homework and rewriting it to make it sound like your own,”} she added. \textit{“Even if no one finds out, you still know what you did.”}

Other metaphors further reinforced this sense of concealment. \( P27_{\text{male, journalist}} \) referred to using GenAI as \textit{“ghostwriting.”} She remarked, \textit{“Ghostwriting is not uncommon in journalism, especially in commercial collaborations. In many Chinese news reports involving corporate clients, the company drafts the content and the newspaper simply publishes it under a journalist’s byline. But it’s never openly discussed. It’s a hidden shame. If we don’t talk about ghostwriting, why would we publicly talk about GenAI?”} In this framing, GenAI’s most powerful capability—text generation—is morally likened to unethical journalistic practices such as plagiarism or covert authorship and is not celebrated as innovation. These comparisons reflect not only individual discomfort, but also the normative environment in which journalists operate: one in which journalistic value is rooted in human creativity, intellectual labor, originality, and textual craftsmanship.

Many participants expressed concern that GenAI could devalue the profession itself, transforming journalism from a human enterprise into something replaceable by machines. As \( P24_{\text{female, journalist}} \) self-mockingly noted, \textit{“We often talk about what professions AI might replace. News reports usually say it will replace low-creativity jobs like data entry. But if even high-literacy, high-creativity journalism can be assisted by GenAI, doesn’t that say something about how little society values our work?”} 

\( P3_{\text{male, editor/manager}} \) shared a widely discussed case in the Chinese news industry that illustrates this anxiety. \( P3_{\text{male, editor/manager}} \) said a prominent Chinese newsroom recently sought to develop its own GenAI system for news writing and asked journalists to contribute training data by submitting ten of their most accomplished past articles. Yet, almost no journalist complied. As \( P3_{\text{male, editor/manager}} \) explained, journalists feared that handing over their best work would not only strip them of control over the copyright of their writing but also allow the system to absorb their style and voice, erasing the individuality of their labor. For many, the idea that their copyrighted articles could be repurposed as raw material for a machine epitomized the risk of being replaced by the very system trained on their work. This moral framing of authorship, as something earned through struggle and talent, helps explain why GenAI remains a taboo topic, even within news organizations that encourage experimentation.

Fear of peer and supervisory judgment also played a critical role. In environments where editorial authority is tied to experience and skill, admitting to AI assistance might be interpreted as a lack of competence or diligence. Junior journalists especially felt that revealing GenAI usage might be seen as laziness or unprofessionalism. \( P27_{\text{male, journalist}} \) reflected, \textit{“My editor is very old-school. If I told him I used AI to structure a piece or just organize interview notes, he’d think I wasn’t taking the job seriously.”} This perception of stigma reinforces secrecy and prevents GenAI from being openly coordinated into collaborative work, even when it could enhance efficiency or creativity.

\section{Findings: Opportunities Toward Future Coordination of GenAI into Collaborative Newswork (RQ3)} \label{opportunity}

Our participants further articulated concrete visions for how GenAI could be embedded into the future of collaborative newswork. Our findings show that journalists anticipate coordinating GenAI in future collaboration will require rethinking organizational priorities (\autoref{rethink1}), fostering a culture of responsible use (\autoref{foster}), and developing better infrastructures (\autoref{tools}) and new workflow (\autoref{create}) that embed GenAI into collaborative practice. 

\subsection{Rethinking Organizational Prioritization: Making Collaboration a Core Design Goal} \label{rethink1}

Toward the end of many interviews, participants began to reflect critically on the organizational priorities that have shaped their newsroom’s GenAI projects. A recurring concern was the observation that internal AI teams and technical leadership tend to prioritize model development and task-specific performance over the collaborative workflows and user interfaces necessary for these tools to function meaningfully in daily editorial practice.

As \( P1_{\text{female, editor/manager}} \), who oversees organizational-level technology and workflow decisions in her newsroom, explained, \textit{“When resources are limited and especially when the government wants to see concrete GenAI system design outcomes, such as a functional prototype system that can be showcased in public or to the government as evidence of progress.”} This mindset, while understandable in a results-driven institutional context like China’s, implicitly relegates collaboration to a downstream concern: something to be added later, once the core AI infrastructure is in place, rather than a foundational element of system design.

To address this imbalance, participants called for a cultural shift in GenAI development: collaboration could be treated as a first-order concern, not an afterthought. They outlined three specific strategies for achieving this transformation: First, involving editors and journalists early in the development process. Rather than viewing users as passive recipients of prebuilt systems, participants argued that journalists could be embedded in the development cycle from the beginning to ensure GenAI tools are grounded in collaborative realities. As \( P18_{\text{male, editor/journalist}} \) put it, \textit{“If no one is thinking about how these tools work in collaboration, how our colleagues or editors might perceive their use, it’s hard to trust that the tools are reliable in a newsroom collaboration context.”}

Second, embed collaboration into the technical design of GenAI systems. Participants emphasized that GenAI platforms could support collaborative features, such as shared editing environments, real-time commenting, and visible histories of GenAI interactions. As \( P8_{\text{male, editor}} \) noted, \textit{“As an editor, I need to know how journalists interact with GenAI. If that visibility were built into the platform, it would help. But without it, we obviously can’t monitor every journalist’s chat history with their GenAI, that would be outrageous and a violation of privacy.”}

Third, expand GenAI evaluation metrics to include collaboration outcomes. Currently, internal GenAI pilot programs and evaluations often focus narrowly on performance: how well the system generates news text, images, or video. But participants argued that such metrics miss a critical dimension: whether GenAI improves or undermines collaborative newswork. As \( P12_{\text{female, journalist}} \) suggested, \textit{“I’m not sure how to measure it, but maybe future evaluations could also include whether these tools actually enhance or hurt teamwork.”}

\subsection{Fostering a Cultural Shift in Newsrooms: Embracing Responsible GenAI Use as Legitimate Newswork} \label{foster}

Several participants called for a broader cultural transformation within their organizations: the normalization and legitimization of responsible GenAI use in journalistic practices. Although GenAI tools are already widely adopted by front-line journalists and editors individually, participants noted that their use remains largely hidden, ad hoc, or even stigmatized. Journalists worry that acknowledging reliance on GenAI might be interpreted as laziness, a lack of editorial rigor, or diminished professional credibility, especially in legacy newsrooms where traditional reporting values continue to dominate.

To address this hesitation, participants emphasized the importance of building a newsroom culture in which responsible GenAI use is not only accepted but also recognized as a legitimate form of editorial labor. \( P20_{\text{male, journalist}} \) remarked, \textit{“We couldn’t pretend everyone is doing everything by hand. That’s not honest, and it discourages people from trying. The reality is, many journalists are already using GenAI. We can’t pretend no one is.”} \( P20_{\text{male, journalist}} \) further explained that if GenAI use is becoming inevitable, then the next step could be to promote responsible use, not prohibit it, particularly given the practical impossibility of policing private GenAI use on personal devices. 

To support this shift, participants urged newsroom leadership to provide clear internal messaging: the use of AI tools, when transparent, accountable, and thoughtfully coordinated, could not be seen as undermining journalistic professionalism. In contrast, such use may reflect a strategic and informed approach to emerging technologies in pursuit of journalistic goals.

Participants also suggested that this cultural shift requires redefining what it means to be a responsible journalist in the GenAI era. Instead of viewing journalistic integrity as the absence of AI involvement, integrity could be evaluated based on the transparency and oversight exercised in using these tools. \( P3_{\text{male, editor/manager}} \) emphasized, \textit{“Responsible GenAI use could be part of our professional standards, not something we hide from our peers or our audience.”}

In participants’ vision, this transformation would also create space for open dialogue and peer learning around GenAI practices. \textit{“Journalists could more freely exchange prompt techniques, usage tips, and editorial guidelines”} (\( P8_{\text{male, editor}} \)), while \textit{“teams could collaboratively reflect on how GenAI use aligns with shared values”} (\( P12_{\text{female, journalist}} \)). 

\subsection{Embedding GenAI into Newsroom Infrastructures for Future Collaboration}  \label{tools}

Participants emphasized the importance of integrating GenAI tools into newsroom technical infrastructures to support future collaborative work. Currently, most journalists rely on external interfaces, such as Qwen Chat, Doubao, or DeepSeek, to experiment with GenAI in their everyday newswork. This results in fragmented workflows and individualized practices that are invisible to other peers. Editors often cannot determine when GenAI was used, what prompts were given, or how much of the final draft was AI-generated, making it difficult to assess originality, editorial appropriateness, AI-like stylistic traces, or factual accuracy. 

In contrast, according to our participants, embedding GenAI functionalities directly into newsrooms' internal content management systems could potentially help transform GenAI from a siloed tool into an auditable and collectively accountable part of journalistic work. According to \( P3_{\text{male, editor/manager}} \), if all interactions, such as prompts, generated drafts and post-editing revisions, were automatically captured and stored on a central platform of the newsroom, editors would be able to trace AI involvement, clarify accountability and make informed editorial decisions in the future. \( P1_{\text{female, editor/manager}} \) agreed that, news organizations already investing in GenAI development can extend that investment to support collaborative visibility and editorial control. 

In this context, the coordination of GenAI is also a way to maintain institutional legitimacy and prepare for potential government audits or public scrutiny regarding how AI is used in sensitive editorial decisions. In some state-run newsrooms, such as CCTV and CGTN, every piece of news could already pass through a centralized editorial platform that records each publication decision \cite{ShenZhou2025Reconstruction,CCF2024PlatformConstruction}. As \( P1_{\text{female, editor/manager}} \) explained,\textit{ “We already have content management systems where every news article is tracked and every editorial decision is logged. So it makes sense to extend that to AI as well, especially in China, where we need to be prepared to justify our editorial decisions at any time, whether to the government or to the public.”}

In \( P1_{\text{female, editor/manager}} \)’s imaginations, this kind of coordination supports risk mitigation at scale. By embedding GenAI into existing editorial infrastructures, media organizations can respond more quickly to reputational, legal, or political concerns if AI-generated content is later questioned. It also creates a shared point of accountability for team members, strengthening collective responsibility and reducing the likelihood that GenAI use remains a hidden, individualized practice outside the editorial chain of command.

\subsection{Creating New Workflow Around GenAI} \label{create}

Several participants proposed creating entirely new workflow that are uniquely enabled by GenAI, workflows that would be difficult, if not impossible, to realize under traditional human-only newswork. These workflow shall reflect a transformation in how content is produced, how collaboration unfolds, and how audiences engage with journalistic output in the age of GenAI.

\subsubsection{Firstly, a common suggestion was the creation of dedicated AI news zones within newsrooms.} These editorial spaces are deliberately separated from core news operations and designed to foster experimentation with GenAI tools. As \( P2_{\text{female, editor/manager}} \) put it,\textit{ “In these AI news zones, audiences know it’s okay to try something new. We’re not pretending it’s traditional reporting.”}

These AI zones were described as low-risk, high-flexibility environments where journalists teams could develop novel formats and experiment with creative applications of GenAI. Participants envisioned a wide range of possibilities that would be difficult to achieve through conventional human-only news production. For instance, some proposed GenAI-powered chatbots that \textit{“allow audiences to converse with digital journalist avatars to better understand news stories” }(\( P22_{\text{male, journalist}} \)). Short-form videos and animations, scripted with GenAI and targeted at younger, mobile-first audiences, were also frequently mentioned, with \( P7_{\text{male, editor}} \) noting that \textit{“there’s already a lot on Douyin, and they’re popular”}. Still others proposed AIGC documentaries composed from archival footage and GenAI-generated narration. \( P11_{\text{female, journalist}} \) explained: \textit{“It’s technically feasible, and we have extensive archives, especially rich local footage that can be used to create documentaries about local cultures.”}

Because these GenAI products are clearly distinct in form and tone from conventional journalism, our participants believe audiences tend to be more tolerant of stylistic experimentation and minor inaccuracies. Editors, in turn, feel freer to iterate on editorial norms. As \( P11_{\text{female, journalist}} \) reflected, \textit{“When people see AI-generated content in these special AI news zones, they don't judge it heavily by the standards of our front page. They expect it to be more innovative.”}

Opening new workflow for GenAI does not eliminate editorial control in our participants’ expectations. As \( P8_{\text{male, editor}} \) emphasized, even content produced in AI news zones could undergo human review, be clearly labeled as AI-generated, and be documented internally. However, the emergence of these formats offers a structured space to explore the creative frontier of GenAI without forcing immediate reform of core editorial systems. \textit{“In these zones,”} \( P8_{\text{male, editor}} \) added, \textit{“journalists and editors can be more creative in how they collaborate with GenAI, without fear of professional judgment or reputational risk.”} 

\add{Despite this enthusiasm for clearly demarcated AI news zones, many participants cautioned that even explicit GenAI labeling could become problematic when AI-generated content appears outside these designated spaces. Several participants (such as \( P11_{\text{female, journalist}} \) and \( P24_{\text{female, journalist}} \)) worried that if a GenAI-labeled product is placed alongside traditional news stories, the label itself may trigger suspicion among readers and colleagues. Rather than reassuring audiences, an AI tag in a conventional news section could lead them to question the authenticity of surrounding content, prompting concerns such as \textit{“If this one is AI-generated, how many others might be as well?”} (\( P24_{\text{female, journalist}} \)) Participants emphasized that, without the protective boundary of AI zones, GenAI labels risk contaminating the credibility of adjacent human-written stories and creating unnecessary verification work. As \( P14_{\text{female, editor/journalist}} \) said, \textit{“Even if you label it, once AI shows up in the regular news pages, people start doubting everything around it.”} For this reason, journalists viewed spatial and categorical separation, not labeling alone, as essential to preserving trust, clarity, and workflow stability in the broader newsroom environment.}

\subsubsection{Secondly, several participants also envisioned a new collaborative newswork form augmented by GenAI.} \( P2_{\text{female, editor/manager}} \) highlighted an experimental system called Fabarta Agents \cite{wjs2025fabarta}, currently being trialed at CCTV and several provincial news televisions as an internal newsroom application. A key feature of the GenAI system is its deliberate emphasis on peer learning and stylistic imitation. Journalists can recall the full corpus of a preferred journalist colleague's past work in the system, using it not only as training material but also as real-time reference during writing. The agent can, for example, suggest opening paragraphs, headlines, or structural outlines based on how a trusted peer might have approached similar stories. As \( P2_{\text{female, editor/manager}} \) explained, \textit{“You can ask the agent to write in the style of someone you respect in the newsroom. It’s like getting mentorship through the GenAI system.”}

Importantly, Fabarta Agents was also designed to address copyright concerns that have accompanied other newsroom GenAI initiatives. According to P2, because the system operates as an internal application, journalists can upload their own past articles without fear that the content will be redistributed outside the organization or used for commercial purposes. In this way, copyright remains under newsroom control, and materials are treated as part of a closed professional resource, available only for internal reference or individual experimentation. P2 emphasized that this feature reassured many journalists who might otherwise resist contributing their work, as it positioned GenAI-assisted stylistic learning as a form of collegial exchange rather than a threat to ownership.

According to \( P2_{\text{female, editor/manager}} \), Fabarta Agents opens up a new pathway for collegial influence and inspiration, where AI becomes a medium for professional emulation and reflection, not merely automation. As \( P2_{\text{female, editor/manager}} \) summarized, \textit{“To me, this GenAI tool essentially creates a safe space for experimentation while also improving our collaborative culture. It shows how technology itself can reshape the very nature of collaboration.” } \add{For \(P8_{\text{male, editor}}\), one major challenge in developing internal AI tools is \textit{“copyright”}. He noted that \textit{“journalists are often worried about how their data might be used or repurposed by the system”}. In his view, a well-designed workflow could help alleviate these concerns by clearly establishing boundaries on how content is stored and reused, and ensuring that individual contributions are protected.}

\section{Discussion}
Our findings suggest that while journalists widely adopt GenAI tools at the individual level, structural and cultural barriers prevent these practices from scaling into sustained teamwork. At the same time, journalists envision productive futures for GenAI in collaboration. These findings offer important design implications for HCI and CSCW researchers seeking to embed GenAI into collaborative work, particularly in high-stakes, institutionalized domains such as journalism. 

This discussion section unpacks three interrelated dimensions. First, it calls for HCI and CSCW research to move beyond a predominant focus on individual interactions with GenAI and instead attend to its role in collaborative newsroom work (\autoref{from}). Second, it rethinks GenAI’s technical essence in contemporary collaboration, as GenAI could help with writing and editing or other creative tasks, thereby affecting how collaboration itself could be organized and legitimated (\autoref{rethink}). Thirdly, drawing from the Chinese newsroom context, we extend these insights to broader cultural and institutional settings (\autoref{global}). Finally, it outlines concrete design and policy implications to support more transparent, accountable, and responsible forms of future coordination of GenAI into collaborative newswork (\autoref{design}).

\subsection{From Individual GenAI Use to Collaborative Work}\label{from}

While recent HCI and CSCW literature has explored the growing role of GenAI in creative and knowledge work, most of this research has focused on individual interactions with AI systems: how users prompt, evaluate, and iteratively shape GenAI outputs to support tasks such as writing, ideation, or coding \cite{palani2024evolving,sun2024generative,lee2025impact,10.1145/3613904.3642114,ma2025speculative}. In contrast, our study contributes to an emerging but underdeveloped area: \add{how individualized GenAI use intersects with collaborative work}, particularly in high-stakes, institutionally-bound domains such as journalism. It also raises forward-looking questions about how organizations can embed GenAI into future collaboration.

Our findings reveal that despite widespread individual experimentation with GenAI usage, newsrooms struggle to transform these fragmented practices into shared norms or team-level workflows. This aligns with long-standing HCI and CSCW insights that new technologies rarely slot in to existing work practices without reconfiguration of roles, norms, and communication structures \cite{moller2024designing,holtgrewe2014new,blomberg2013reflections,geiger2021labor,hu2022distance,ma2024my}. Coordinating GenAI into collaboration in our cases is a socio-technical challenge, encompassing both the design of collaborative interfaces and the need to ensure accountability, visibility, interpretability, and trust among newsroom actors.

In the technical aspect, our study suggests that \add{the opacity of GenAI interactions—both in how prompts are crafted by journalists and in how outputs are generated by the model—renders shared understanding and joint decision-making difficult, a dynamic that participants referred to as “dual black-boxing.”} This dual black-boxing of human prompting and machine generation creates asymmetries in editor–journalist collaboration, echoing prior HCI and CSCW concerns about collaboration breakdowns in common ground when collaborators cannot access or reconstruct one another’s interactions with complex systems \cite{carroll2006awareness,ackerman2013sharing}. 

Furthermore, from the cultural aspect, individual GenAI use tends to remain private, informal, and even stigmatized, which Goffman might call backstage behavior \cite{goffman1949presentation}. Goffman \cite{goffman1949presentation} distinguishes between “frontstage” behavior, where individuals perform in alignment with institutional norms and expectations, and “backstage” behavior, which is concealed from public view and often deviates from formal roles. In our study, journalists’ use of GenAI largely occupies this backstage space: it is done alone, without disclosure, and often with ambivalence about its legitimacy. This concealment has important implications for how collaborative GenAI practices do or do not take shape. Because GenAI use is hidden, it resists normalization: it cannot be scrutinized, iterated upon, or collectively stabilized. As a result, organizations struggle to build legitimate, durable, and visible practices around AI collaboration. 

\add{The challenges and opportunities of moving from individualized GenAI use to collaborative newsroom practices also hinge on how different roles perceive the risks and benefits of coordination. For front-line journalists, GenAI provides pragmatic relief under heavy workloads, yet coordination also risks exposing invisible writing labor to new forms of evaluation. Their ambivalence reflects a tension between the practical benefits of GenAI and the reputational concerns associated with making its use visible. Editors encounter the dual black box from the opposite direction: without access to either prompting processes or model reasoning, they face heightened uncertainty in verification and accountability. Although shared standards could reduce ambiguity and improve review processes, editors also worry that formal coordination might institutionalize practices that could undermine editorial authority or public credibility.} \add{Managers approach GenAI from an institutional perspective shaped by efficiency, compliance, and organizational reputation. While coordinated GenAI systems are often envisioned as potential infrastructures for improving workflow consistency, managers remain wary of the political and ethical risks that organizational endorsement could entail. Informal use localizes risk to individual journalists, whereas coordinated use redistributes liability upward, increasing exposure at the organizational level. Together, these differentiated stances explain why individualized GenAI use persists even amid calls for coordination: the move from private experimentation to collective practice entails uneven risk redistribution across roles, making coordination a site of negotiation rather than a straightforward technical transition.} 

\subsection{GenAI as a Collaborative Technology in Future Newsroom} \label{rethink}

Unlike earlier collaborative technologies that supported newsroom work primarily in peripheral ways, GenAI is entering news production at points much closer to the substantive core of journalistic tasks. Rather than replacing existing collaborative norms outright, this shift introduces new pressures for renegotiating roles, workflows, and professional legitimacy. Prior HCI, CSCW, and organizational research has long demonstrated that technologies such as groupware, workflow systems, and enterprise platforms reshape collaboration by redistributing tasks, restructuring coordination, and altering authority \cite{rama2006survey,poltrock1998cscw,grudin1999cscw}. These earlier systems, however, generally operated around the margins of knowledge work: they facilitated communication, scheduling, or documentation but rarely engaged directly with the content that defined professional outputs \cite{rama2006survey,knearem2023exploring,xiao2025institutionalizing}. 

\add{GenAI introduces similar dynamics of negotiation, yet with a different locus of impact}. Instead of remaining in the background, GenAI more directly interacts with the content of journalistic work by drafting headlines, summarizing interviews, or generating preliminary article versions, as discussed in \autoref{Culture} and \autoref{foster}. This integration into content creation surface tensions more immediately tied to authorship, editorial authority, and responsibility. In this sense, GenAI prompts journalists and editors to articulate what counts as legitimate professional practice and how accountability should be shared.

Our findings find that journalists rarely disclosed their use of GenAI to colleagues or editors, not only due to coordination challenges such as tracing AI contributions, but also because of concerns about skill erosion, replacement, and copyright—all themes that resonate with prior HCI work on participation and secret AI use \cite{suresh2024participation,zhang2025secret,waters2025shadow}. \add{These anxieties illustrate that GenAI’s integration into newsroom collaboration intersects with existing professional norms and identities, rather than acting as a singular, disruptive force.}

For GenAI to contribute constructively to collaborative newswork, practices may need to evolve beyond hidden, individualized use toward more visible and shared organizational norms. As shown in \autoref{motivations} and \autoref{Culture}, secrecy can undermine trust, obscure accountability, and introduce uncertainty into team-based decision-making. These dynamics arise because its current use intersects with long-standing professional expectations about originality, expertise, and editorial judgment. Thus, GenAI draws attention to fundamental questions about how journalistic work is defined and evaluated, and how collaborative practices might adapt when automated tools participate more directly in core writing tasks.

Journalism scholars and practitioners may therefore benefit from shifting attention from private experimentation toward the cultural and organizational conditions that enable more transparent and principled uses of GenAI. Future HCI and CSCW research could examine how the introduction of GenAI into collaborative newswork reframes professional identity, and how teams and organizations negotiate boundaries between human judgment and computational assistance. More broadly, embedding GenAI into collaborative practice involves the familiar challenge of aligning new technological capabilities with collective standards of fairness, independence, and accuracy, rather than relying on isolated or concealed forms of use.

\add{In many ways, GenAI extends long-standing themes in CSCW and organizational studies: new technologies enter existing ecologies of work, redistribute tasks, and prompt negotiations over authority, coordination, and accountability \cite{valentine2017flash,xiao2025might,herrmann2023keeping}. As with earlier collaborative systems, GenAI’s adoption unfolds through situated practice, shaped by organizational history, local norms, and the social meanings attached to technological use. At the same time, GenAI differs in where these negotiations occur. Whereas prior systems primarily reshaped the supporting infrastructures of collaboration, GenAI operates at the level of content creation itself, making questions of authorship, voice, and professional judgment more salient \cite{lewis2025generative,munoriyarwa2025generative}. This shift reorientate familiar CSCW concerns toward domains of work that were previously less technologically mediated. In other words, GenAI follows established patterns of how collaborative technologies become domesticated in organizations, yet it situates these dynamics closer to journalistic labor, where professional identity and accountability are most tightly held in writing, editing and creativity.}

\subsection{Extending the Chinese Insights to Other Cultural and Institutional Contexts} \label{global}

\add{While our study is situated in the Chinese newsroom context, the dynamics we identify extend beyond China and offer broader implications for theorizing GenAI-supported collaboration across diverse media systems. Our findings do not merely describe a local phenomenon; instead, they surface core socio-technical patterns that can inform how media organizations in other countries understand and govern GenAI in collaborative newswork, echoing prior research on China-based cases that has generated insights applicable to global contexts \cite{xiao2025artists,liu2023neoliberal,kuai2022ai}.}

\add{First, China provides a uniquely revealing site for studying GenAI integration because newsrooms operate under strong institutional directives to demonstrate technological modernization and political responsiveness \cite{xiao2025might,kuai2024unravelling,kuai2025artificial}. Participants described how AI system-building was often oriented toward external legitimacy, for instance, satisfying government expectations for digital transformation, rather than addressing the everyday collaborative practices of journalism. This top-down pressure creates symbolic, rather than operational, GenAI integration, generating a misalignment between organizational incentives and newsroom-level collaborative needs. Our findings therefore highlight a mechanism: when GenAI adoption is driven by external legitimacy pressures rather than newsroom collaborative needs, collaboration may suffers. Although political oversight varies across countries, similar tensions may emerge in media systems influenced by market competition, corporate mandates, or public scrutiny. This suggests that our Chinese case offers a conceptual lens to diagnose how external institutional pressures shape the collaborative form of GenAI coordination in other contexts.}

\add{Second, China’s hierarchical newsroom structures reveal how GenAI becomes entangled with power asymmetries in collaborative work. Editors and managers bear political and reputational risk, making them more cautious about GenAI-assisted workflows, while junior journalists privately use GenAI to cope with workload demands \cite{kuai2025navigating}. This pattern reflects a broader insight: GenAI adoption is fundamentally shaped by local distributions of authority, responsibility, and risk. Even in less centralized media systems, power asymmetries, between senior and junior journalists, editors and freelancers, or center and periphery newsrooms, can strongly influence who adopts GenAI, who discloses its use, and who bears responsibility for errors. Our findings therefore offer a transferable framework for examining how GenAI adoption interacts with organizational hierarchies across diverse professional cultures.}

\add{Third, China's cultural norms around experimentation, public accountability, and media professionalism shape how journalists disclose or conceal GenAI use. In our study, the backstage nature of GenAI adoption reflects not only coordination challenges but also deeper cultural anxieties about maintaining credibility in a tightly regulated informational environment. Journalists feared that admitting GenAI use might be read as violating professional identity, weakening editorial rigor, or risking political misinterpretation. These findings extend Goffman's frontstage/backstage framework \cite{goffman1949presentation} by illustrating how backstage practices are shaped not only by personal identity work but also by broader sociopolitical conditions. This insight is generative for cross-cultural studies: transparency and disclosure norms around GenAI are likely to vary globally and will co-evolve with local journalistic values, professional identities, and cultural expectations about automation.}

\add{These insights show that the Chinese case is a theoretically productive site that foregrounds a triad of forces, i.e., political oversight, organizational hierarchy, and cultural norms, that influence GenAI's role in collaboration. These forces are present to varying degrees across other media ecosystems. The Chinese case thereby provides both analytical tools and empirical reference points that other researchers can adapt when investigating GenAI adoption in their own institutional and cultural contexts.}

\subsection{Design and Policy Implication} \label{design}

We call for bringing GenAI use into newsrooms out of the shadows and embedding it into collaborative, transparent, and accountable practices that align with both everyday workflows and institutional values. \add{Building on our findings, we argue that better coordinating GenAI into collaboration in newsrooms requires interventions at two levels: (1) bottom-up design that equips everyday users with tools for transparency, coordination, and shared control, and (2) top-down policy structures that institutionalize accountability and align GenAI use with organizational values. This dual perspective reflects longstanding HCI guidance that successful human–AI systems could integrate interaction-level and institution-level interventions. \add{Table~\ref{tab:design_policy} summarizes the key implications that we elaborate in the following two subsections, distinguishing bottom-up design considerations from top-down policy needs.}}

\begin{table*}[t]
\centering
\caption{\add{Summary of Key Design and Policy Implications for Collaborative GenAI Coordination in Newsrooms}}
\label{tab:design_policy}
\begin{tabular}{p{0.23\linewidth} p{0.32\linewidth} p{0.32\linewidth}}
\toprule
\textbf{Key Challenge Area} & \textbf{Bottom-Up Design Implications} & \textbf{Top-Down Policy Implications} \\
\midrule

\textbf{Transparency and Traceability (marked by 6.1.2, 6.2, 7.2)} 
& 
Make human–AI contributions visible across workflows; provide prompt histories, revision paths, and provenance indicators. 
& 
Require traceable logging of GenAI involvement, retention policies, and disclosure guidelines aligned with editorial standards. \\
\midrule

\textbf{Collaborative Workflows (marked by 6.1.1)} 
& 
Design shared GenAI workspaces, multi-user prompt libraries, and co-editing environments that reflect newsroom teamwork.
& 
Define organizational expectations for collaborative GenAI use; incentivize early front-line journalist participation in AI tool development and evaluation. \\
\midrule

\textbf{Institutional Norms as Interface Affordances (marked by 6.2)} 
& 
Embed newsroom values into interface-level prompts: e.g., reminders for attribution, accuracy checks, and ethical flags.
& 
Establish formal norms for responsible GenAI use; build institutional structures for transparency, auditing, and consistent editorial review. \\
\midrule

\textbf{Learning and Trust-Building (marked by 5.2, 6.2)} 
& 
Support gradual user learning with features enabling peer knowledge sharing and cross-role visibility.
& 
Create shared prompt libraries, peer-learning forums, and organizational training programs supporting GenAI literacy. \\
\midrule

\textbf{Accountability Structures (marked by 6.1.2, 6.1.3)} 
& 
Design mechanisms to clarify responsibility for AI-assisted content, especially in multi-party editing tasks.
& 
Develop clear accountability frameworks specifying editorial responsibility, escalation paths, and how AI errors are handled. \\
\midrule

\textbf{Scalability and Governance (marked by 6.1.2)} 
& 
Build interaction-level affordances that scale across teams (e.g., configurable transparency levels for editors vs. journalists).
& 
Integrate GenAI into editorial infrastructures; prepare for regulatory demands and reputational risk management in high-stakes reporting. \\
\bottomrule
\end{tabular}
\end{table*}

\subsubsection{Designing for Collaborative GenAI Integration}

\add{From a bottom-up perspective, effective collaborative GenAI coordination could begin with interaction-level design that supports journalists’ real practices.} 

\add{First, design for social traceability to counter the “double black box.” In our study, GenAI use was often invisible on two levels: editors and colleagues rarely knew when or how journalists were using GenAI, and journalists themselves often lacked clarity on how the model produced its outputs. This double opacity not only undermines accountability but also weakens trust in collaborative processes. Design could therefore make both human and AI contributions more visible across the editorial workflow. For instance, interfaces could visually differentiate between human-generated and AI-generated text, display prompt histories, and show revision paths. This design stance aligns with Human–AI Interaction guidelines that call for visible system reasoning and meaningful traceability by translating them into concrete mechanisms for editorial collaboration, clarifying how human-AI actions interact with human judgment across the workflow \cite{amershi2019guidelines}.}

Second, design for shared editorial workflow. Current GenAI tools primarily operate in a single-user paradigm, where one journalist privately interacts with the system. Yet our participants increasingly envisioned GenAI as a shared infrastructure woven into teamwork. To realize this vision, design could enable collaborative use: shared GenAI workspaces, prompt libraries editable by multiple users, and co-edit outputs that record collective contributions. \add{These ideas resonate with existing HCI design recommendations encouraging systems to support collaborative use cases and enable cross-role coordination in AI-augmented work \cite{amershi2019guidelines,deng2023investigating,xiao2025might}.}

Third, turning institutional norms into interface-level affordances. Journalists sought tools that would help them uphold values of transparency, attribution, and editorial integrity. Systems can support this by prompting disclosure of GenAI involvement, suggesting attribution formats consistent with newsroom policies, or flagging content that might raise ethical concerns. Integrating these practices into the interface itself would allow journalists to enact shared values in everyday work, helping normalize GenAI as a legitimate, collectively governed part of newswork. \add{This design direction follows Human–AI guidelines such as “match relevant social norms" and "mitigate social biases" which are especially important in high-stakes domains like journalism \cite{li2025actions,amershi2019guidelines}.}

\subsubsection{Toward Collaborative and Accountable GenAI Policies}

\add{While bottom-up design interventions can scaffold everyday collaborative use, they could be complemented by top-down organizational policies that institutionalize responsible GenAI use at scale. Without structural support, interaction-level transparency or coordination mechanisms cannot fully address issues of accountability, risk, and uneven power dynamics identified in our findings.}

First, to meaningfully support coordination of GenAI into collaborative newswork, newsrooms could treat collaboration not as a downstream usability concern, but as a strategic design priority. Institutional policies could incentivize the early involvement of front-line journalists in AI development processes and require that GenAI evaluation metrics extend beyond technical benchmarks like accuracy. Instead, assessments could include collaborative usability, editorial interpretability, and shared accountability across team members. \add{Such alignment between organizational policies and interface-level design mirrors the Human–AI Interaction guideline encouraging alignment between AI capabilities and organizational expectations \cite{li2025makes,murire2024artificial,rakova2021responsible}.}

Second, newsrooms could establish clearer norms and incentives to institutionalize the responsible use of GenAI. Our findings indicate that informal, often stigmatized use of GenAI persists in part because of the absence of formal guidance and supportive peer practices. Policies could not treat GenAI use as a compliance problem, but as a cultural opportunity to scaffold shared learning and responsible exploration. Practices such as shared prompt libraries, peer-learning forums, and transparent audit trails can foster a culture of open dialogue, where GenAI use is not hidden but collaboratively refined and improved. \add{This echoes the HAI guideline to support user learning across time, which emphasizes that GenAI competence could grow within a social and organizational context \cite{amershi2019guidelines,murire2024artificial,li2025makes}.}

Third, for GenAI to be governable and sustainable in the long term, it could be integrated into the newsroom’s editorial newswork. We emphasize that GenAI adoption and coordination could occur at the infrastructural level. Policies could require traceable logging of GenAI involvement in content production, guidelines for data retention, and clear delineation of editorial responsibility when AI-generated content is used. These mechanisms not only support internal accountability and collaboration, but also prepare news organizations to meet external demands, such as regulatory compliance, reputational risk management, and public scrutiny over AI usage in high-stakes reporting environments. \add{By pairing these top-down governance frameworks with bottom-up design changes, newsrooms can learn from HCI scholars and move toward a socio-technical ecosystem that supports transparent, accountable, and trustworthy GenAI-enabled collaboration \cite{aitamurto2019hci}.}

\section{Limitations and Future Work}
\add{This study provides early and exploratory insight into the emerging role of GenAI in collaborative newswork. As an exploratory qualitative investigation, our findings illuminate emergent patterns in how journalists personally experiment with GenAI and how these practices create both tensions and opportunities for collaboration. However, this exploratory nature also brings several limitations that point toward important directions for future research.} 

\add{First, our sample is limited to Chinese newsrooms, which operate within a distinct media system shaped by strong institutional norms and political constraints. While many of the structural and cultural dynamics we observed may resonate globally, our findings could not be interpreted as universally representative. Future work could examine how GenAI adoption varies across different media ecosystems, including commercial, public service, and nonprofit news organizations in democratic contexts. Comparative studies could further explore how political systems, regulatory environments, and organizational cultures shape the institutionalization of GenAI-supported collaboration. Examining related sectors such as education, law, or healthcare may also reveal contrasting pathways for incorporating GenAI into high-stakes collaborative practices.}

\add{Second, our recruitment strategy and study design may have favored journalists who were already using GenAI directly and were relatively open or positive toward experimenting with it. Because many participants were recruited through personal networks and snowball sampling, those who are skeptical, disengaged, or more deeply concerned about GenAI may be underrepresented. This limitation is compounded by the reliance on self-reported accounts, which can introduce biases in how participants frame their own practices and attitudes. Future work could seek to incorporate more diverse perspectives, including journalists who resist, avoid, or strategically withhold GenAI use, and could combine interviews with direct observation, trace data, or system logs to mitigate self-report bias. }

\add{Third, while our findings speculate on possible future workflows and organizational coordination, we lack deployment-based or behavioral data to demonstrate whether the proposed collaborative mechanisms would indeed enhance multi-party trust, coordination, or efficiency in practice. Our analysis identifies potential workflows and cultural transformations, but empirical evidence from real-world deployment or longitudinal observation is needed to validate whether these workflows truly support collaborative accountability, reduce invisible labor, or shift newsroom culture. Future work could therefore explore the design, prototyping, and in situ deployment of collaborative GenAI systems, studying how such systems are appropriated, resisted, or reshaped over time. }

\add{Finally, although our study focused primarily on editorial staff, particularly journalists and editors, collaborative GenAI use in news organizations may extend beyond editorial roles. Designers, developers, product managers, and data teams increasingly contribute to the integration of GenAI in newsroom infrastructures. Future work could adopt a more holistic organizational lens to trace how cross-functional teams negotiate GenAI integration, how institutional priorities shape design and deployment decisions, and how organizational boundaries and power asymmetries influence collaborative uptake.}

\section{Conclusion}
Through 27 interviews with senior editors, newsroom managers and front-line journalists in China, this study examined how journalists navigate the coordination of GenAI into collaborative newswork. We identified a clear gap between individual experimentation with GenAI and the structural and cultural mechanisms needed to support sustained, legitimate, and team-based use. While journalists expressed curiosity and even enthusiasm about GenAI's potential, their current practices remain fragmented, private, and stigmatized. Our paper contributes to HCI and CSCW by foregrounding the sociotechnical barriers that prevent GenAI from becoming a collaborative infrastructure and pointing out future opportunities to coordinate GenAI into collaborative work. 

\bibliographystyle{ACM-Reference-Format}
\bibliography{references}

\appendix

\section{Interview Protocol and Demographic Questions} \label{protocol}

\add{This appendix provides the full interview protocol and the demographic questions used to contextualize participants' backgrounds and collaborative roles. The protocol followed a semi-structured, three-phase design, allowing researchers to probe deeply into participants’ concrete practices, expectations, and critical reflections on GenAI in collaborative newswork.}

\subsection{Demographic Questions}

\add{At the beginning of each interview, participants were asked a brief set of demographic questions to contextualize their professional background and position within the newsroom. These questions were optional and used solely to support analysis of GenAI practices across different roles and experience levels.}

\begin{itemize}
    \item \add{\textbf{Gender:} How do you identify your gender?}
    \item \add{\textbf{Age:} What is your age?}
    \item \add{\textbf{Years in Journalism:} How many years have you worked in journalism or media production?}
    \item \add{\textbf{Current Position:} What is your role in the newsroom?}
    \item \add{\textbf{Newsroom Unit or Department:} Which desk, team, or section are you part of? (e.g., society, politics, investigations, video)}
    \item \add{\textbf{Primary Collaborative Partners:} With whom do you most often collaborate in daily newswork?}
    \item \add{\textbf{Experience with GenAI Tools:}}
    \begin{itemize}
        \item \add{Have you used GenAI tools in your reporting or editing work?}
        \item \add{If yes, for which tasks?}
        \item \add{Frequency of use: (Several times per day / Daily / Several times per week / Occasionally)}
    \end{itemize}
    \item \add{\textbf{Involvement in GenAI Decision-Making:}}
    \begin{itemize}
        \item \add{Are you involved in discussions or planning related to GenAI adoption in your newsroom?}
    \end{itemize}
\end{itemize}

\subsection{Interview Protocol}

\add{The interviews were semi-structured and organized into three main phases. Follow-up probing questions were used as appropriate to deepen or clarify participants’ accounts.}

\subsubsection{Phase 1: Concrete Experiences with GenAI in Newswork}

\begin{itemize}
    \item \add{Can you describe specific moments when you used GenAI in your daily reporting or editing work?}
    \item \add{What types of GenAI tools or platforms did you use, and how did you access them (e.g., internal tools, public LLMs)?}
    \item \add{At which points of the reporting or editing process did you most commonly use GenAI (e.g., idea generation, background research, drafting, editing, verification)?}
    \item \add{How did GenAI use unfold in collaborative contexts, for example with editors or teammates?}
    \item \add{Were there moments when you used GenAI together with colleagues, or shared AI-generated content with others? How did they respond?}
    \item \add{Can you recall situations where GenAI changed how responsibility or workload was distributed within a team?}
    \item \add{In what ways did GenAI fit or fail to fit into your existing newsroom routines?}
    \item \add{Did GenAI use ever create uncertainty or confusion in collaborative work (e.g., accountability, authorship, reliability)?}
    \item \add{How did your organization communicate or coordinate early-stage GenAI practices? Were there any informal norms or expectations?}
    \item \add{How did GenAI shape your interactions with editors, colleagues, or other teams?}
    \item \add{Have you ever chosen not to use GenAI in a situation where you could have? Why?}
    \item \add{Have you witnessed colleagues using GenAI in ways that affected your own work—positively or negatively?}
    \item \add{How did GenAI affect the pace, accuracy, or style of your reporting or editing tasks?}
    \item \add{Did GenAI introduce any new forms of invisible work (e.g., checking, correcting, validating outputs)?}
\end{itemize}

\subsubsection{Phase 2: Expectations for Collaborative GenAI Integration}

\begin{itemize}
    \item \add{In your ideal scenario, how would GenAI support your team’s collaborative work?}
    \item \add{What kinds of assistance would you want GenAI to provide for teamwork or cross-role coordination?}
    \item \add{What would make a GenAI tool feel reliable, trustworthy, or compatible with collaborative newswork?}
    \item \add{How would you envision GenAI interacting with editors, senior managers, or fellow journalists?}
    \item \add{If GenAI could understand your team’s workflow, what aspects could it pay attention to?}
    \item \add{What would an “ideal” GenAI system know about your newsroom’s editorial standards or organizational values?}
    \item \add{How proactively could GenAI participate in teamwork?}
    \item \add{What kinds of transparency would you need from GenAI to feel comfortable using it collaboratively?}
    \item \add{How could GenAI tools display provenance, e.g., where ideas, edits, or facts came from, when supporting collaborative tasks?}
    \item \add{Do you imagine different roles in the newsroom using GenAI differently? How could tools adapt to these differences?}
    \item \add{What would an optimal division of labor between humans and GenAI look like in reporting, editing, and publishing?}
    \item \add{If you could redesign your current workflow with GenAI in mind, what changes would you make?}
    \item \add{How could GenAI support cross-team workflows?}
    \item \add{Are there any risks or unintended consequences that an ideal collaborative GenAI system could help prevent?}
    \item \add{What kinds of training or onboarding would help your team use GenAI more effectively together?}
    \item \add{How could GenAI tools negotiate ethical concerns?}
    \item \add{In what ways could GenAI strengthen teamwork culture, e.g., coordination, communication, shared understanding?}
\end{itemize}

\subsubsection{Phase 3: Gaps Between Ideals and Current Practices}

\begin{itemize}
    \item \add{Why do you think GenAI has not yet become a regular part of your team’s workflow?}
    \item \add{What challenges, breakdowns, or tensions arise when colleagues use—or do not use—GenAI?}
    \item \add{What forms of uncertainty, hesitation, or resistance have you observed in your team?}
    \item \add{What organizational or technological changes would make GenAI more compatible with collaborative newsroom practices?}
    \item \add{What concerns or risks do you foresee in deeper GenAI integration?}

    \item \add{Can you recall any specific instances when GenAI created confusion, misalignment, or extra coordination work?}
    \item \add{Have you experienced situations where teammates used GenAI but did not disclose it? How did this affect collaboration?}
    \item \add{What makes it difficult for colleagues to trust GenAI—or to trust each other when using GenAI?}
    \item \add{How do power dynamics (e.g., junior journalists vs. editor) affect willingness to use or share GenAI-generated content?}

    \item \add{How does the lack of clear guidelines or protocols influence GenAI use in your team?}
    \item \add{Do you think the current newsroom workflow or CMS makes it hard to incorporate GenAI into collaboration?}
    \item \add{Are there concerns about accountability when GenAI is used in reporting or editing?}
    \item \add{What kinds of errors or inaccuracies have you seen from GenAI that created extra work for your team?}
    \item \add{Are there legal, political, or reputational risks that discourage GenAI adoption in your newsroom?}

    \item \add{Are there differences between how willing different roles (e.g., editors, junior journalists, managers) are to use GenAI?}
    \item \add{Have organizational priorities or incentives made it harder (or easier) to integrate GenAI into team workflows?}

    \item \add{What prevents GenAI from supporting more complex collaborative tasks?}
    \item \add{If you imagine your ideal workflow again, what are the biggest gaps between that vision and your current reality?}
\end{itemize}

\section*{Notes}
\begin{itemize}
    \item \add{The protocol was used flexibly and adapted to participants with different newsroom roles.}
\end{itemize}

\end{document}